\documentclass[final,5p,times,twocolumn,authoryear]{elsarticle}

\journal{Medical Image Analysis}

\usepackage{amssymb}
\usepackage{latexsym}
\usepackage{mathtools}
\usepackage{multicol}
\usepackage[textsize=small]{todonotes}
\usepackage{booktabs}       
\usepackage{multirow}

\usepackage{url}
\usepackage{xcolor}
\usepackage{hyperref}
\usepackage{cleveref}

\newcommand{\textblue}[1]{\textcolor{black}{#1}}

\definecolor{newcolor}{rgb}{.8,.349,.1}

\journal{Medical Image Analysis}

\newcounter{suppfigure}
\newenvironment{suppfigure}[1][]{%
    \figure[#1]%
    \refstepcounter{suppfigure}%
}{%
    \endfigure
}

\newenvironment{suppfigure*}[1][]{%
    \begin{figure*}[#1]%
    \refstepcounter{suppfigure}%
}{%
    \end{figure*}%
}

\newcommand{\suppcaption}[1]{%
    \caption{#1}%
}

\begin{document}


\begin{frontmatter}

\title{Towards contrast-agnostic soft segmentation of the spinal cord}%
\tnotetext[]{\textsuperscript{\textdagger}these authors contributed equally to this work}

\author[1]{Sandrine Bédard\textsuperscript{\textdagger}}
\ead{sandrine.bedard@polymtl.ca}
\author[1,2]{Enamundram  Naga Karthik\corref{cor1}\textsuperscript{\textdagger}}
\cortext[cor1]{Corresponding author: naga-karthik.enamundram@polytmtl.ca}
\author[3,4,5]{Charidimos Tsagkas}
\author[6]{Emanuele Pravatà}
\author[3,4,5]{Cristina Granziera}
\author[7]{Andrew Smith}
\author[8]{Kenneth Arnold Weber II}
\author[1,2,9,10]{Julien Cohen-Adad}
\ead{jcohen@polymtl.ca}

\address[1]{NeuroPoly Lab, Institute of Biomedical Engineering, Polytechnique Montréal, Montréal, Québec, Canada}
\address[2]{Mila - Québec Artificial Intelligence Institute, Montréal, Québec, Canada}
\address[3]{Translational Imaging in Neurology (ThINK) Basel, Department of Biomedical Engineering, Faculty of Medicine, University Hospital Basel and University of Basel, Basel, Switzerland}
\address[4]{Department of Neurology, University Hospital Basel, Basel, Switzerland}
\address[5]{Research Center for Clinical Neuroimmunology and Neuroscience Basel (RC2NB), University Hospital Basel and University of Basel, Basel, Switzerland}
\address[6]{Neuroradiology Department, Neurocenter of Southern Switzerland, Ospedale Regionale di Lugano, Lugano, Switzerland}
\address[7]{Department of Physical Medicine and Rehabilitation Physical Therapy Program, University of Colorado School of Medicine, Aurora, Colorado, USA}
\address[8]{Division of Pain Medicine, Department of Anesthesiology, Perioperative, and Pain Medicine, Stanford University School of Medicine, Stanford, California, USA}
\address[9]{Functional Neuroimaging Unit, CRIUGM, University of Montreal, Montreal, Québec, Canada}
\address[10]{Centre de recherche du CHU Sainte-Justine, Université de Montréal, Montréal, Québec, Canada}



\begin{abstract}
Spinal cord segmentation is clinically relevant and is notably used to compute spinal cord cross-sectional area (CSA) for the diagnosis and monitoring of cord compression or neurodegenerative diseases such as multiple sclerosis. While several semi and automatic methods exist, one key limitation remains: the segmentation depends on the MRI contrast, resulting in different CSA across contrasts. This is partly due to the varying appearance of the boundary between the spinal cord and the cerebrospinal fluid that depends on the sequence and acquisition parameters. This contrast-sensitive CSA adds variability in multi-center studies where protocols can vary, reducing the sensitivity to detect subtle atrophies. Moreover, existing methods enhance the CSA variability by training one model per contrast, while also producing binary masks that do not account for partial volume effects. 
In this work, we present a deep learning-based method that produces soft segmentations of the spinal cord that are stable across MRI contrasts. 
Using the Spine Generic Public Database of healthy participants ($\text{n}=267$; $\text{contrasts}=6$), we first generated participant-wise soft ground truth (GT) by averaging the binary segmentations across all 6 contrasts. These soft GT, along with aggressive data augmentation and a regression-based loss function, were then used to train a U-Net model for spinal cord segmentation.
We evaluated our model against state-of-the-art methods and performed ablation studies involving different GT mask types, loss functions, contrast-specific models \text{and domain generalization methods}. 
Our results show that using the soft average segmentations along with a regression loss function reduces CSA variability ($p < 0.05$, Wilcoxon signed-rank test). 
The proposed spinal cord segmentation model generalizes better than the state-of-the-art contrast-specific methods amongst unseen datasets, vendors, contrasts, and pathologies (compression, lesions), while accounting for partial volume effects.

\end{abstract}


\end{frontmatter}




\section{Introduction}

Spinal cord segmentation is clinically relevant, notably to compute cross-sectional area (CSA) for the diagnosis and monitoring of atrophy in multiple sclerosis (MS)~\citep{Barkhof1997-qo}, spinal cord injury (SCI) \citep{TROLLE2023103372}, and in characterizing spinal cord compression~\citep{Martin2018-qt}. 
While several approaches for semi-automatic and automatic segmentation of the spinal cord have been introduced~\citep{De_Leener2016-dt,Gros2019-uq,Horsfield2010-xl}, they all suffer from the same limitation: the output segmentation depends on the MRI contrast and acquisition parameters of the input image ~\citep{Gros2019-uq}. For instance, the CSA measured from T2-weighted (T2w) images is approximately 8\% higher than that measured from T1-weighted (T1w) images ~\citep{Cohen-Adad2021-su, Kim2015-jz}. This contrast-dependency is partly due to the varying appearance of the boundary between the spinal cord and the cerebrospinal fluid because of differences in MR properties (e.g., relaxation times, spin density, flow). Different acquisition parameters and pulse sequences produce different contrast and sharpness of the spinal cord boundary, which consequently affect the output of the segmentation methods, whether they are manual, semi-automatic or automatic. The contrast-sensitive CSA also adds variability in multi-site studies, thereby reducing the sensitivity to detect subtle atrophies~\citep{Bautin2021-qe}.

One way to mitigate the impact of MRI contrast on metrics derived from the segmentation is to compute the CSA on various contrasts and estimate a scaling factor based on the contrast type as done by~\citep{Cohen-Adad2021-su} for T1w and T2w contrasts and MRI vendors. However, the scaling factors themselves are highly dependent on the MRI vendor, pulse sequence and imaging parameters, thus limiting their application to other studies. 

Recent work addressed the contrast dependency of automatic segmentation in terms of model performance. \cite{Gros2019-uq} trained separate deep learning models for each contrast. However, the ground truth (GT) masks used for training were generated using a combination of automatic PropSeg \citep{De_Leener2014-tk} and manual corrections, which were done separately for each contrast. As a result, the GT masks for each contrast were already biased, resulting in models that robustly segmented the spinal cord, but produced different CSA across contrasts (e.g., higher T2w CSA than T1w CSA). 
SynthSeg \citep{Billot2020-ch, Billot2020-vl, Billot2023-ua}, a deep learning-based method primarily used for the segmentation of brain MRI scans of any contrast or resolution, leveraged GT label maps during training to synthetically generate brain images of various contrasts. While SynthSeg provides segmentations that are inherently agnostic to the input image contrast, it relies on a domain randomization strategy, where parameters such as orientation, resolution, and contrast are randomly sampled from a uniform distribution to synthetically generate the training scans. However, this requires a large number of GT segmentations, which are difficult to obtain for spinal cord scans that often include various structures such as the vertebrae, the spinal canal, intervertebral discs, nerve rootlets, surrounding muscles, the lungs and the heart. 

Generalization to unseen domains is a paramount objective for deep learning algorithms. Domains can be defined as sets of images acquired from different sites and scanners, images consisting of contrasts other than those in training, or even images containing pathologies (i.e., lesions) when trained on healthy images. 
Domain generalization methods in the literature treat this as a domain shift problem at the fundamental level, where each contrast, for example, is seen as a different but related domain with minor differences in their marginal distributions \citep{Guan2022-ak}. Such methods propose to use domain adaptation techniques to \textit{transfer} the differences between the source and target domains, either by mapping both domains to a shared latent space \citep{kamnitsas2017unsupervised, dou2019pnp, ganin2016domain} or by generatively adapting the source to target domains by image-to-image translation methods \citep{hoffman2018cycada, chen2019synergistic}. 
Expanding further into the concept of learning domain-invariant features, regularization, as a means of creating a representative feature space consisting of various domains has also been explored \citep{Dou2019DG, Li2020DomainGF, zhang2022semi}. 
Other related works include meta-learning for adapting model for few-shot weakly-supervised segmentation tasks \citep{Oliveira2022DG} and adversarial training for increasing the diversity of the training data \citep{zhang2023domain}.
While unsupervised methods in domain adaptation alleviate the need for labeled training data, such methods still need re-training on each subsequent target domain \citep{Bateson2022-ih}, which is impractical. 

\textblue{Data augmentation-based domain generalization methods aim to model the domain shifts via a series of transformations applied to the input images at the source domains during training.}
For instance, \cite{zhao2019data} proposed a learning model for spatial transformations to synthesize additional labeled examples for one-shot segmentation in brain MRI scans. \cite{Ouyang2023CIS} used causality-driven data augmentation specifically targeting domain shifts and acquisition shifts, while \cite{su2023slaug} proposed a location-scale augmentation using Bezier transformations, both in the context of single-source domain generalization. \cite{Ling2020GDL} showed that simply relying on sequential stack of data augmentation transforms based on image quality, appearance and spatial configuration, results in good generalization to unseen domains.  


In addition to the contrast-dependent issues discussed above, one of the main limitations of traditional segmentation methods is that they produce binary (hard) segmentation masks, which do not account for partial volume effects  \citep{Billot2020-vl, Chaves2021-jw}. Partial volume effect is characterized by mixing of signals from different tissues within the same voxel, resulting in averaged intensities which are not representative of any of the underlying tissues. Binary masks do not provide calibrated output probabilities for the partial volume information of the tissue. With soft labels, the segmentation is encoded with continuous values between 0 and 1 and can therefore encode partial volume information, while resulting in better generalization ~\citep{Gros2021-ms}, faster learning~\citep{Muller2019-ys}, and increase the precision of voxel-based morphometry or CSA measurements \citep{Lemay2022-me}. 

\subsection{Contributions}

In this work, we present a convolutional neural network (CNN) model for the automatic soft segmentation of spinal cord across various contrasts. Our model reduces the variability in CSA across contrasts and generalizes to spinal cord images of unseen contrasts and pathologies. Our original contributions are as follows: 
\begin{enumerate}
    \item We introduce a new pipeline for generating a unique, soft GT that represents the segmentations across various MRI contrasts. 
    \item Contrary to \citep{Gros2021-ms} where the softness was obtained implicitly after data augmentation, we propose to apply the data-augmentation transforms directly on the soft GT masks and train a contrast-agnostic SoftSeg model for spinal cord segmentation. 
    \item We show that the proposed model reduces variability morphometric measures (i.e., produces stable soft segmentations across contrasts) and shows significant improvement over prior work using contrast-specific models and domain-generalization methods.
\end{enumerate}

The model is open-source and the code for pre-processing/training/inference can be found in the following GitHub release\footnote{https://github.com/sct-pipeline/contrast-agnostic-softseg-spinalcord/tree/v2.0}. \textblue{It is also integrated into the Spinal Cord Toolbox and available in v6.2\footnote{https://github.com/spinalcordtoolbox/spinalcordtoolbox/releases/tag/6.2} and higher}.
The rest of the paper is structured as follows: in Section \ref{sec:mat-and-meth}, we describe the training dataset, preprocessing pipeline, the training and evaluation protocols. In Section \ref{sec:results}, we show the results from the various validation experiments and comparisons with baselines and state-of-the-art methods. Different features and perspectives of the proposed spinal cord segmentation model are discussed in Section \ref{sec:discussion}, followed by the conclusion in Section \ref{sec:conclusion}.

\section{Materials and Methods}
\label{sec:mat-and-meth}

\subsection{Dataset}
\label{subsec:dataset}
We used the Spine Generic Public Database\footnote{\href{https://github.com/spine-generic/data-multi-subject}{https://github.com/spine-generic/data-multi-subject}} (Multi-Subject) \citep{Cohen-Adad2021-su} consisting of 267 healthy participants scanned across multiple MRI vendors (Siemens, GE and Philips) and scanner models. Each participant has a 3D T1-weighted MPRAGE (T1w) at 1mm isotropic resolution, 3D T2w at 0.8mm isotropic resolution, 2D T2*w axial at $0.5\times0.5\times3$ mm (multi-echo GRE), 3D axial gradient-echo with (MT-on) and without (GRE-T1w, with a shorter TR and a higher flip angle compared to the MT-on scan) magnetization transfer pulse at $0.9\times 0.9 \times 5$ mm, and an axial diffusion-weighted scan motion corrected and averaged across diffusion directions at $0.9\times0.9\times5$ mm. This multi-contrast dataset was chosen because it is publicly available, and it includes a large variety of MR contrasts that are popular in the MR community. Participants with missing contrasts or excessive artifacts were excluded from our experiments ($n=24$ out of 267).

The final dataset included 243 participants with 6 contrasts each, resulting in 1458 3D volumes in total. These were split according to 60/20/20 train/validation/test splits, resulting in 145 participants (870 volumes) for training, 49 participants (294 volumes) for validation and 49 participants (294 volumes) for testing.

\subsection{Data preprocessing for ground truth generation}
\label{subsec:prepro}
\begin{figure*}[h!]
\centering
\includegraphics[width=0.95\linewidth]{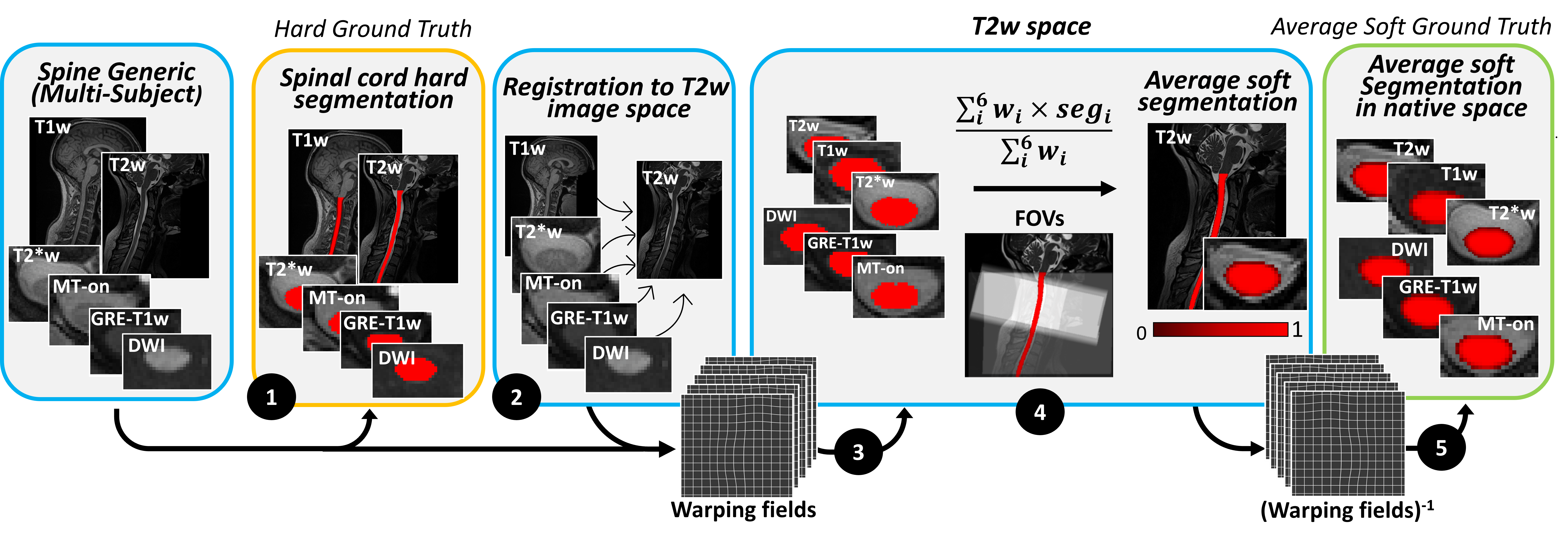}
\caption{Preprocessing pipeline for soft average segmentations ground truth. (1) Automatic hard spinal cord segmentation using \texttt{sct\_deepseg\_sc} \& manual corrections; (2) Registration to T2w space; (3) Applying each contrast's warping field to bring the segmentation masks to the T2w space; \textblue{(4) Weighted averaging of segmentations according to each contrast FOV (represented by white rectangles) to create a unique soft GT mask} (5) Applying inverse warping fields to bring the unique soft GT to the native space of each contrast.}
\label{fig:preprocess}
\end{figure*}

To eliminate the differences in CSA within the GT across contrasts, we used a unique segmentation averaged over all contrasts as the GT for training. 
\textblue{Our objective here was to obtain the soft segmentation resulting from each contrast-specific hard segmentation}. Figure ~\ref{fig:preprocess} shows an overview of the procedure for generating the GT using SCT~\cite{De_Leener2017-gq}. The GT soft segmentations are generated by averaging 6 different contrasts (T1w, T2w, T2*w, MT-on, GRE-T1w and DWI). For each participant and contrast, the spinal cord was segmented using SCT’s \texttt{sct\_deepseg\_sc} to generate a binary segmentation. Manual corrections were made when significant segmentation errors (i.e., leaking and under-segmentation) were observed in SCT's quality control report.  \textblue{Since \texttt{sct\_deepsg\_sc} (DeepSeg2D) is considered the state-of-the-art for spinal cord segmentation and creating GT from scratch is highly time consuming, we obtained an initial batch of segmentations followed by manual quality control.}

All images and binary segmentations were registered to the T2w image space as it has the highest resolution (0.8 mm isotropic). The registration was done using SCT's \texttt{sct\_register\_multimodal} center-of-mass algorithm. It consists of a slice-by-slice alignment of the center of mass of the input and target segmentations (rotation and translation in x and y directions). The registration was performed in 10 iterations with a gradient step of 0.5. The segmentations of all 6 contrasts were then averaged within the T2w space to obtain a unique average soft segmentation (ranging from 0 to 1). The average of the segmentations was weighted according to the field-of-view (FoV) of each contrast. 
\textblue{More precisely, we created a mask of the FoV of each contrast by dilating the spinal cord segmentation in the axial plane to get the complete superior-inferior coverage. Then, we registered these FoV masks to the common T2w space (see Figure \ref{fig:preprocess}, step 4), such that the contrasts with overlapping FoVs are weighted more in the unified soft GT.} Then, the averaged soft segmentation was brought back to each contrast's native space by applying the inverse warping field and using linear interpolation. This step was important because having the GT in the native space of each contrast eliminates biases due to various resolutions and fields-of-view during training. The original images along with the soft GT masks (both in their native space), were then used during training.

The vertebral levels were automatically labeled using SCT's \texttt{sct\_label\_vertebrae} command on the T1w and T2w images. Quality control was done using the \texttt{sct\_qc} command and when necessary, labels were manually created at the posterior tip of each intervertebral disc. Since the contrasts T2*w, MT-on,  GRE-T1w and DWI are axial acquisitions with thick slices, the manual or automatic labeling of the discs is not reliable. To generate the vertebral labels for those contrasts, we warped the T2w intervertebral discs to each contrast's native space using the generated warping field in the previous step. Finally, we computed spinal cord CSA on the soft average segmentation GT and the binary segmentations averaged over the C2-C3 vertebral levels using SCT's \texttt{sct\_process\_segmentation}. 

\subsection{Training Protocol}
\label{subsec:training}

In this section, we describe the (online) preprocessing, data augmentation, our proposed model and the training strategy used for the contrast agnostic segmentation of the spinal cord. 

\subsubsection{Preprocessing}

All data were resampled to $1 \textrm{mm}$ isotropic resolution and re-oriented to right-posterior-inferior (RPI) before training. The images and the GT labels were resampled using spline interpolation and linear interpolation, respectively. \textblue{The median shape of all the images in the training set after resampling was $192 \times 230 \times 106$}.
As a consequence of having images with different orientations (3D, axial) and fields-of-view (cervical, cervico-thoracic, thoracic), we found center cropping to be extremely useful. Notably, the images were heavily cropped in the R-L and A-P directions to keep the spinal cord at focus while the S-I direction was left uncropped. The final patch size for center-cropping was set to $64 \times 192 \times 320$.

\subsubsection{Data Augmentation}

Given the heterogeneity across contrasts, heavy data augmentation was crucial for the performance of the model. 
All data augmentation transforms are random, applied with a pre-defined probability and called in the following order: 
affine transformation with spline and linear interpolation for images and labels respectively $(\text{p}=0.9)$ \textblue{and the rotation, scaling and translation parameters ranging between $[-20, 20]$, $[-0.2, 0.2]$, and $[-0.1, 0.1]$, respectively}, 
elastic deformation ($\text{p}=0.5$) \textblue{by sampling a grid of random offsets within $[25, 35]$ and Gaussian smoothing the grid with the standard deviation (STD) between $[3.5, 5.5]$}, 
simulation of low resolution ($\text{p}=0.25$) \textblue{with a downsampling and upsampling factors sampled uniformly from $[0.5, 1.0]$}, gamma correction ($\text{p}=0.5$) \textblue{with magnitude between $[0.5, 3.0]$, where $1.0$ gives the original image and smaller/larger value makes image lighter/darker, respectively}, 
bias field adjustment ($\text{p}=0.3$) \textblue{with the range of random coefficients between $[0.0, 0.5]$}, 
Gaussian noise addition ($\text{p}=0.1$) \textblue{with mean $0.0$ and the STD spread uniformly between $[0.0, 0.1]$},
Gaussian smoothing ($\text{p}=0.3$) \textblue{with the STD of the smoothing kernel ranging from $[0.0, 2.0]$ for all axes}, 
intensity scaling ($\text{p}=0.15$) \textblue{by multiplying in the range $[-0.25, 1.0]$}, 
random mirroring ($\text{p}=0.3$) (\textblue{across all axes )}. 
Lastly, all images were normalized (independently) using $z$-score normalization by subtracting the mean intensity and dividing by standard deviation intensity. These augmentation transforms are readily implemented in MONAI \citep{Jorge_Cardoso2022-ik}.


\subsubsection{Model Architecture}

Given the popularity of the nnUNet \citep{Isensee2021-ja}, we used the same architectural template found in nnUNet's \texttt{3d\_fullres} model\footnote{\href{https://github.com/MIC-DKFZ/dynamic-network-architectures/blob/main/dynamic_network_architectures/architectures/unet.py}{https://github.com/MIC-DKFZ/dynamic-network-architectures/unet.py}}. Each layer in the encoder and decoder contains two blocks, each consisting of a convolutional layer, instance normalization \citep{Ulyanov2016-aw} and leakyReLU non-linearity \citep{Maas2013-gc}. Strided convolutions are used for downsampling while transposed convolutions are used for upsampling. Additionally, the network is trained with deep supervision \citep{Dou2017-ia}, where auxiliary losses from the feature maps at each upsampling resolution are added to the final loss. This allows for the gradients to be injected deeper into the network, thus facilitating the training of all layers. The encoder made up of 5 layers, starting with 32 feature maps at the initial layer and ending with 320 feature maps at the bottleneck (i.e. $32 \to 64 \to 128 \to 256 \to 320$).


Unlike nnUNet, which uses softmax activation on the logits, we followed the SoftSeg approach \citep{Gros2021-ms} and used normalized ReLU (NormReLU) as the final activation function. This choice is made from the observation that activation functions like sigmoid and softmax have a polarizing effect that undesirably shorten the range of soft values that carry valuable partial volume information at the boundaries. NormReLU simply normalizes the output of the ReLU activation using the maximum value, which is given by:
\begin{equation}
\text{NormReLU}(x) = 
\begin{cases}
\frac{\text{ReLU}(x)}{\max \text{ReLU}(x)} & \text{if max ReLU}(x) \neq 0\\
0 & \text{otherwise}
\end{cases}
\end{equation}

\begin{figure}[htbp!]
\centering
\includegraphics[width=0.475\textwidth]{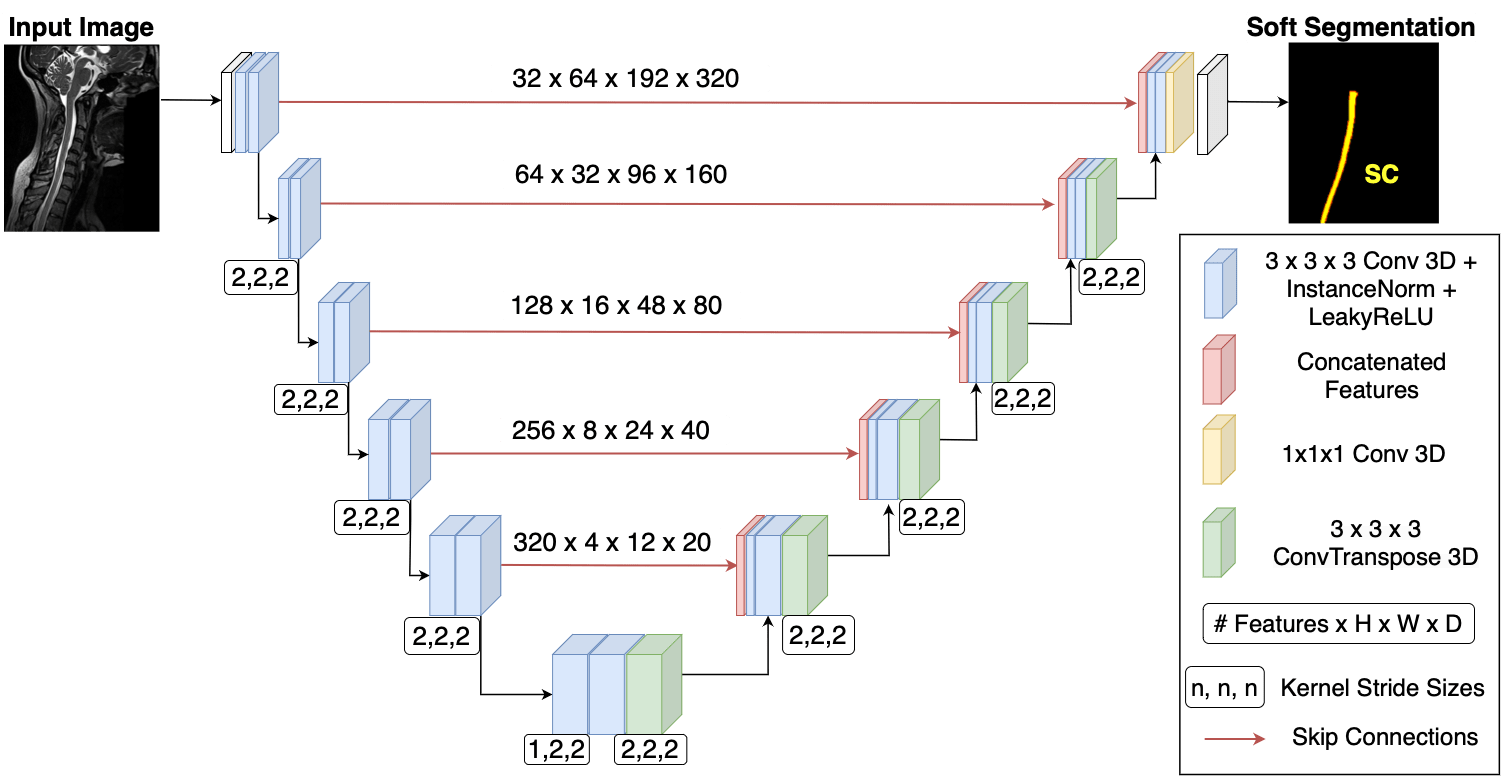}
\caption{Architecture of the proposed SoftSeg model.}
\label{fig:model}
\end{figure}

This offers the advantage of preserving the useful properties of the ReLU activation function while ensuring that the predictions are normalized within the range of 0 and 1. A graphical representation of the model architecture is shown in Figure \ref{fig:model}.

\subsubsection{Loss Function}
\label{subsubsec:loss-func}

An issue with the commonly used DiceLoss \citep{Milletari2016-xj} is that it yields segmentation masks with sharp edges \citep{Deng2018-vk}. It has also been shown the optimizing with soft Dice leads to volumetric biases (due to under-/over-segmentation) with high inherent uncertainty \citep{Bertels2020-sl}. More specifically, for our contrast agnostic segmentation problem, DiceLoss does not drive the model towards optimizing for accurate segmentations at the boundary between the spinal cord and the cerebrospinal fluid, once it obtains a \textit{good enough} segmentation of the spinal cord. 
This was also observed by \cite{Jia2019-lc} in the context of left atrium segmentation in cardiac images.
Therefore, the inability of DiceLoss to adapt its behaviour upon reaching closer to convergence (i.e., failing to distinguish between the spinal cord / cerebrospinal fluid boundary), is the primary factor contributing towards the differences in CSA across contrasts. 
\textblue{Although not as severe as just the DiceLoss, the} same applies to other loss functions such as the Focal \citep{Lin2017-jw}, Tversky \citep{Salehi2017-df}, \textblue{and Dice-Cross Entropy (DiceCE)} losses. 

As suggested in \cite{Gros2021-ms}, we considered the segmentation task as a pixel-wise regression problem (instead of classification) and trained our model with adaptive wing loss \citep{Wang2019-jb}. An immediate advantage is that a regression objective produces outputs with proper calibration while allowing soft outputs lying in $\left[0-1\right]$. Originally proposed in the context of alignment of facial landmarks via heatmap regression, there are two reasons why adaptive wing loss is a suitable candidate for obtaining soft segmentations. 
\textit{First}, in heatmap regression, the model regresses against the GT heatmap generated by plotting Gaussian distributions centered at each facial landmark. The mode of the Gaussian (i.e. the landmark) and the pixels in its immediate neighbourhood are considered as foreground, while the rest is background. In our case, this presents a similar class imbalance problem where the pixels at the spinal cord / cerebrospinal fluid boundary are outnumbered by pixels at the core of the spinal cord. 
\textit{Second}, loss functions assigning equal weights to all pixels during training (such as DiceLoss) do not result in accurate predictions at the boundaries.
Moreover, pertaining to medical image segmentation, adaptive logarithmic losses have been shown to converge faster and mitigate class imbalance \citep{Kaul2021-ph,Gros2021-ms}.

The loss function is defined as follows: 
\begin{equation}
    \text{AWing}(y, \hat{y}) = 
    \begin{cases}
        \omega \ln \left( 1 + \left| \frac{(y - \hat{y})}{\epsilon} \right|^{(\alpha - y)} \right) & 
        \text{if} \left| (y - \hat{y}) \right| < \theta \\
        A \left|(y - \hat{y}) \right| - C &  
        \text{otherwise},
    \end{cases}
\end{equation}
where $y$ and $\hat{y}$ correspond to GT and the predicted labels, and $\omega$, $\epsilon$, $\theta$, $\alpha$ are the hyperparameters. 
Briefly, the piece-wise loss function has non-linear and linear parts. The former ensures that error between the prediction and GT smaller than $\theta$ have a larger influence in the loss function (via larger gradients during backpropagation), while the latter makes the loss function behave like the mean-squared error loss with equal weights to all voxels. The definitions of the adaptive factor $A$, the constant term $C$ and the hyperparameters can be found in Section 4.2 of \cite{Wang2019-jb}. The following hyperparameter values $\omega=8.0$, $\epsilon=1.0$, $\theta = 0.5$, $\alpha=2.1$ were set for training. We also experimented with $\omega=12.0$, $\epsilon=0.5$ (i.e. larger $\omega$ and smaller $\epsilon$) as suggested in their original work, but did not observe substantial improvement in performance. 

\subsubsection{Hyperparameters \& Training Details}

We used the Adam optimizer \citep{Kingma2014-it} with a learning rate of $0.001$ and a cosine annealing scheduler. The model was trained for a maximum of 200 epochs, and the batch size was set to 2. The patch size for training and sliding window inference was set to $64 \times 192 \times 320$, same as the center cropping size. All the models were trained using the MONAI \citep{Jorge_Cardoso2022-ik} and PyTorch Lightning\footnote{\href{https://lightning.ai}{https://lightning.ai}} frameworks on a single 48 GB NVIDIA A6000 GPU.

\subsection{Evaluation Protocol}
\label{subsec:eval}
In this section, we describe the evaluation protocol to assess the model's performance. 

\subsubsection{Evaluation Metrics}
To quantitatively evaluate the segmentation accuracy, we computed the Dice coefficient (on the binarized predictions thresholded at $0.5$), average surface distance (ASD), and relative volume error (RVE) for each contrast across all test participants. 
To assess the variability of CSA across contrasts, we computed CSA averaged over C2-C3 vertebral levels of the cervical spinal cord on all the test set predictions for each evaluated model. The following metrics were used for quantitative evaluation: 
\begin{enumerate}
    \item \textit{STD CSA}: The standard deviation (STD) of CSA across contrasts for each participant to assess CSA variability,
    \item \textit{Absolute CSA Error}: The absolute error between the CSA of GT segmentation and the prediction for each participant.
\end{enumerate}
Mathematically, the absolute CSA error $\epsilon$ is given by:
\begin{equation}
\epsilon = \; \mid y_{\text{CSA}} - \hat{y}_{\text{CSA}} \mid,
\end{equation}
where $y_{\text{CSA}}$ corresponds to the CSA of the GT segmentation mask and $\hat{y}_{\text{CSA}}$ to the CSA of the prediction averaged at C2-C3 vertebral levels.
The metrics are computed in $\text{mm}^2$ and the lower the STD and absolute CSA errors the better is the model, with the underlying assumption being that one participant should have the same spinal cord CSA across contrasts. 

\subsubsection{Baselines}

We evaluated our model against 3 baselines, each with a different training strategy as described below.

\textbf{Soft vs. Hard ground truth:} \; To assess the impact of the type of GT mask used for training and its consequential effect on CSA variability, we trained two models: one with the soft GT generated using the procedure described in Section \ref{subsec:prepro}, and the second model with the contrast-specific hard GT generated using \texttt{sct\_deepseg\_sc} and manually corrected as required.

\textbf{Single model for all contrasts vs. Contrast-specific models:} \; 
\textblue{Using the soft GT masks}, we trained one model per contrast (i.e. 6 models) and compared it against a single model trained on all 6 contrasts. This experiment is useful in understanding whether a single model, exposed to all contrasts during training, is capable of mitigating the CSA bias.

\textbf{DiceCE loss vs. Adaptive wing loss:} \; As mentioned in Section \ref{subsubsec:loss-func}, optimizing only for the Dice coefficient is insufficient for accurate segmentations at the spinal cord / cerebrospinal fluid boundary. For empirically validating this hypothesis, we treated the loss function as a hyperparameter and trained \textblue{two models with:  (i) Dice-Cross entropy loss, and (ii) adaptive wing loss (the proposed model)} while using both the soft and hard GT segmentations.

Among all baselines, the performance of the models were evaluated in terms of the STD CSA and the absolute CSA errors across all contrasts for each participant in the test set. 
\textblue{Except for the ablation comparing the loss functions, all other models were trained with the adaptive wing loss.}

\subsubsection{Comparison with the state of the art (SOTA)}

We also compared our model's performance with a few SOTA methods, adapting it for spinal cord segmentation wherever necessary.

\textbf{PropSeg}: \; \texttt{PropSeg} \citep{De_Leener2014-tk} is based on the iterative propagation of deformable models for spinal cord segmentation. The algorithm consists of three steps: (i) an initialization step for detecting and orientating the position of the spinal cord using a circular Hough transform, (ii) a propagation step that initializes a deformable model for its propagation along the spinal cord, and (iii), a refinement step for robust and accurate segmentation of the spinal cord. 

\textbf{DeepSeg}: \; \texttt{DeepSeg} \citep{Gros2019-uq}, implemented in SCT as \texttt{sct\_deepseg\_sc}, features a two-stage process: (i), the spinal cord centerline is detected using a 2D CNN with dilated convolutions, and (ii), the cord is segmented along the centerline using a 2D or 3D CNN with standard convolutions.  
This model was trained on 'real world' retrospective data from 30 sites including both healthy participants and pathological patients. 
Images were acquired from various vendors (Siemens, GE, Philips) and included 4 contrasts (T1w, T2w,  T2*w and DWI) with a variety of image resolutions and fields-of-view (axial and sagittal). 
Because of its robustness to multi-site data, \texttt{DeepSeg} is an appropriate benchmark method.

\textbf{nnUNet}: \; The nnUNet \textblue{framework} \citep{Isensee2021-ja} is the SOTA in various segmentation tasks across several challenges. We used the latest version of nnUNet (i.e. \texttt{nnUNetv2}) and train both 2D and 3D variants with the default, self-configured parameters on a single fold for 1000 epochs using all contrasts together and soft GT segmentations binarized using a threshold of $0.5$. This was done because nnUNet does not yet support training with soft GT labels. 

\textblue{
\textbf{SoftSeg}: \; SoftSeg \citep{Gros2021-ms} showed that by skipping the binarization step after data augmentation, one can obtain the soft labels 'for free' and training on these soft GT results in better generalization and calibrated models. Contrary to our approach of creating soft labels by averaging the segmentations of multiple input contrasts \textit{and} applying the data augmentation transforms directly on the soft labels, SoftSeg started with hard labels and trained on the soft labels obtained implicitly after data augmentation. 
}

\textblue{
\textbf{BigAug}: \; \texttt{BigAug} \citep{Ling2020GDL}, is a data augmentation-based domain generalization approach, that applied a series of 9 stacked augmentation transforms based on image quality, appearance and spatial configuration to model domain shifts. While \citep{Ling2020GDL} reported generalization across sites/scanners only within a single contrast (T2w), we adapted their method to compare generalization across different contrasts. Specifically, \texttt{BigAug} was trained on a collection of all 6 contrasts with hard GT labels using Dice loss and evaluated on the basis of STD CSA, absolute CSA error and generalization to unseen contrasts.
}

\textblue{
\textbf{SynthSeg}: \; SynthSeg \citep{Billot2023-ua} is the SOTA method for contrast-agnostic segmentation of brain MRI scans. As it could not be used out-of-the-box for spinal cord segmentation, we re-trained SynthSeg using the segmentation labels for the cord, cerebrospinal fluid, vertebrae, and  intervertebral discs.
The output segmentations from SynthSeg were not comparable to the rest of the methods, hence we report its results in the supplementary material along with its training details.
}

\subsubsection{Generalization to Unseen Data}
As described in Section \ref{subsec:dataset}, the Spine Generic Public Database \citep{Cohen-Adad2021-su} consists of healthy participants only. To evaluate our model's ability to generalize to \textit{real world} clinical data, we tested our model on three datasets of patients presenting various spinal cord pathologies, contrasts and/or on fields-of-view unseen during training.

\textbf{Traumatic Spinal Cord Injury (\texttt{sci-t2w})}:\; 
This dataset consists of axial thoraco-lumbar T2w images of 80 patients with chronic traumatic spinal cord injury from the University of Colorado Anschutz Medical Campus. Acquisition was performed using MRI systems from 2 vendors (Siemens: $n = 16$, GE: $n = 63$) with 2 different field strengths (3T: $n = 17$, 1.5T: $n = 62$) and image resolutions ranging between $\{0.31 - 0.78\} \times \{0.31 - 0.78\} \times \{3-6\}$ $\textrm{mm}^3$. The challenge for the model is to be able to segment the spinal cord, in the presence of spinal cord compression, broken vertebrae and hyperintense lesions (likely edema).  

\textbf{Multiple Sclerosis (\texttt{ms-mp2rage})}: \; This dataset consists of sagittal MP2RAGE "UNI" images ($1 \times 1 \times 1$ mm resolution) of 103 healthy controls and 180 multiple sclerosis patients with visible lesions from the University of Basel acquired on a Siemens MRI scanner. The challenge for the segmentation model here is that the MP2RAGE contrast is unseen during training, and that the hypointense lesions can lead to under-segmentation mainly due to the similar signal intensity as the surrounding cerebrospinal fluid. 

\textbf{Cervical Radiculopathy (\texttt{radiculopathy-epi})}: \; This dataset consists of resting state axial gradient-echo echo-planar-imaging (GRE-EPI) images ($0.89 \times 0.89 \times 5$ mm resolution) of 24 participants with cervical radiculopathy and 28 age- and sex-matched healthy controls from Stanford University acquired on a GE MRI scanner. This dataset was acquired in the context of a resting state functional MRI experiment, and consists of 245 volumes that were motion corrected and averaged. Cervical radiculopathy is characterized by degenerative changes to the cervical spine, which can compress the spinal nerve roots and compromise the normal anatomy of the spinal cord, and the T2*w EPI images that can lead to strong image distortions and signal dropout making the segmentation difficult. 

\textblue{Lastly, we also tested our model on the FLAIR contrast. A few qualitative segmentations are shown in the supplementary material Section S2.
}

\subsubsection{Ablations on the number of contrasts}
As six contrasts could be considered more than what are typically acquired in MRI examinations, we performed two more experiments ablating the number of contrasts in the preprocessing and training stages, and evaluated its downstream effect on the reduction of the CSA variability across all 6 contrasts. Starting with $n=2$ contrasts (T1w and T2w), we followed the same preprocessing pipeline described in Section \ref{subsec:prepro} and trained a model on the soft masks generated by averaging the T1w and T2w contrasts together. The same experiment was repeated for $n=4$ contrasts (T1w, T2w, DWI, T2*w). The results are reported in the supplementary material Section S3.

\section{Results}
\label{sec:results}

In this section, we present the results from our proposed contrast-agnostic spinal cord segmentation model (Section \ref{subsec:contrast-agnostic-seg}) and evaluate them against the baselines (Section \ref{subsec:baselines-comp}) and the existing SOTA methods (Section \ref{subsec:sota-comp}). Then, we show the generalization capabilities of our model on unseen, out-of-distribution data (Section \ref{subsec:generalization}). Lastly, in Section \ref{subsec:inference-time}, we compare the CPU inference times between various methods.

In all the plots in the following sections, the proposed model is denoted by \texttt{soft\_all}, meaning that the model was trained with a soft GT averaged from the individual segmentations of each of the 6 contrasts \textit{and} adaptive wing loss was used as the loss function.

\subsection{Contrast-agnostic spinal cord segmentation}
\label{subsec:contrast-agnostic-seg}

Table \ref{tab:ca-metrics} shows the quantitative results for the proposed contrast-agnostic spinal cord segmentation model \textblue{\texttt{soft\_all}}. For each contrast, we present the mean $\pm$ standard deviation across test participants for Dice coefficients, relative volume errors (in \%), and average surface distances. While the Dice coefficients are consistent across all contrasts, we note a slight under-segmentation in the case of MT-on and DWI contrasts (reflected by the negative RVE) and an over-segmentation for the T1w, T2w, T2*w and GRE-T1w contrasts.

\setlength{\tabcolsep}{8pt}
\begin{table}[htbp!]
    \centering
    \caption{Quantitative results for spinal cord segmentation across contrasts on the test set (49 participants) \textblue{for our \texttt{soft\_all} model}. RVE stands for Relative Volume Error and ASD stands for Average Surface Distance. }
    \resizebox{.5\textwidth}{!}{
    \begin{tabular}{l c c c}
        \toprule
        Contrasts & Dice ($\uparrow$) & RVE \%  & ASD ($\downarrow$) \\
        \cmidrule{2-4}
                  & Opt. value: 1     & Opt. value: 0         & Opt. value: 0 \\
        \midrule
        T1w     & $0.96 \pm 0.02$   & $1.74 \pm 3.38$   & $0.08 \pm 0.25$ \\
        T2w     & $0.96 \pm 0.01$   & $1.89 \pm 2.35$   & $0.01 \pm 0.07$ \\
        T2*w  & $0.96 \pm 0.01$   & $0.56 \pm 2.94$   & $0.01 \pm 0.01$ \\
        MT-on    & $0.96 \pm 0.02$   & $-0.59 \pm 2.88$  & $0.01 \pm 0.03$ \\
        GRE-T1w& $0.95 \pm 0.02$   & $0.99 \pm 5.58$   & $0.04 \pm 0.09$ \\
        DWI     & $0.96 \pm 0.02$   & $-1.04 \pm 3.89$  & $0.00 \pm 0.00$ \\
        \bottomrule
    \end{tabular}
    }
    \label{tab:ca-metrics}
\end{table} 

\begin{figure}[t]
\centering
\includegraphics[width=.475\textwidth]{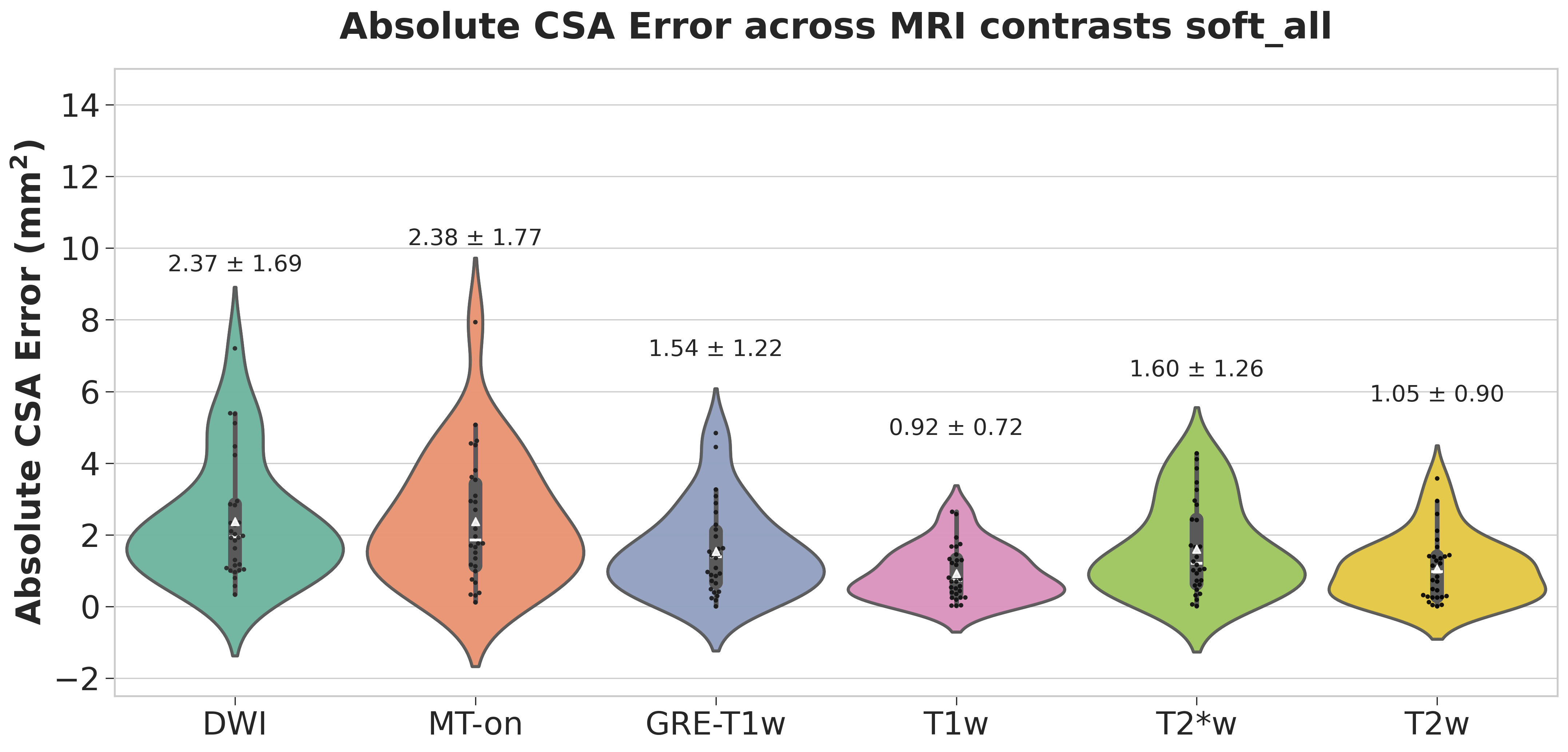}
\caption{Absolute CSA error between the predictions and GT across each contrast for the proposed model. Scatter plots within each violin represent the individual CSA errors for all participants in the test set. White triangle marker shows the mean CSA error across participants.}
\label{fig:soft-all-abs-csa-error-per-contrast}
\end{figure}

Figure \ref{fig:soft-all-abs-csa-error-per-contrast} shows the violin plot with absolute CSA error between the predictions and the GT across 6 contrasts (the lower the better). The mean CSA error is \textblue{less than} 2 $\text{mm}^2$ across all contrasts, which is encouraging given that 2 $\text{mm}^2$ represents only two pixels at an axial resolution of $1 \times 1$ mm.

\begin{figure}[t]
\centering
\includegraphics[width=.475\textwidth]{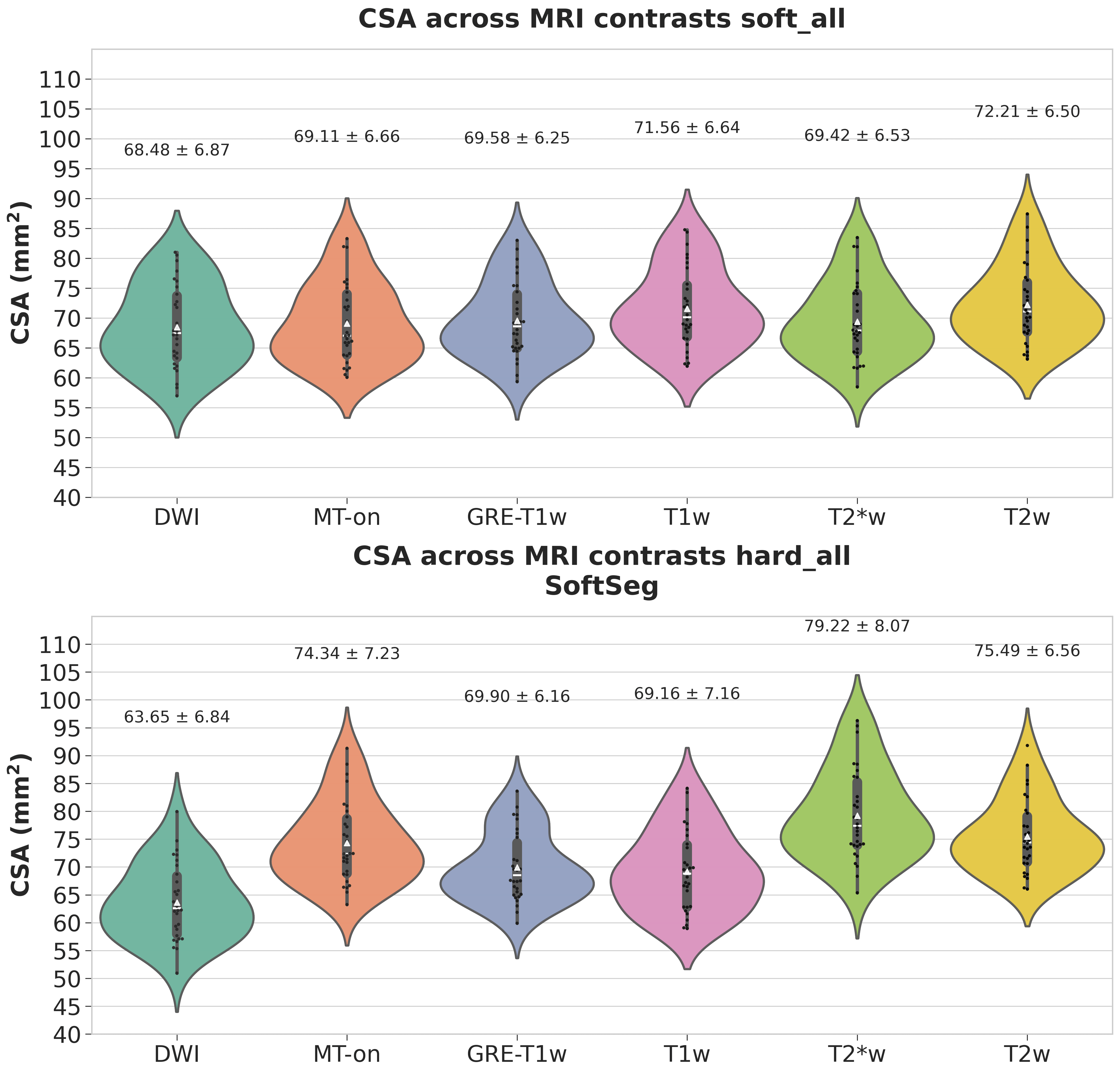}
\caption{Effect of GT segmentation type (soft vs. hard) on CSA across contrasts. White triangle marker shows the mean CSA across participants.}
\label{fig:soft-vs-hard-comparison}
\end{figure}

Figure \ref{fig:soft-vs-hard-comparison} shows the comparison of CSA across contrasts between two models trained with soft GT (top panel) and hard GT (bottom panel). Training with soft GT resulted in similar CSA across all contrasts, while the hard GT training resulted in drastically uneven distributions of CSA. 
\textblue{Note that training with contrast-specific hard GT masks with adaptive wing loss is precisely the training strategy used in SoftSeg \citep{Gros2021-ms}, hence we denote this model as \texttt{hard\_all\_SoftSeg}}. 
We conducted a one-way paired ANOVA on CSA across contrasts for both \texttt{soft\_all} and \textblue{\texttt{hard\_all\_SoftSeg}} models. The MRI contrast had a significant effect on CSA for both methods ($p <0.05$). A follow-up posthoc analysis (two-sided Bonferroni-corrected non-parametric Wilcoxon signed-rank test) revealed that for \texttt{soft\_all}, T2w / T2*w, T2*w / T1w, T1w / GRE-T1w, T2w / GRE-T1w, T2w / MT-on, T2w / DWI, T1w / MT-on, and T1w / DWI pairs of contrast showed a significant difference between CSA ($p <0.05$). While for \textblue{\texttt{hard\_all\_SoftSeg}}, all pairs of contrasts show significantly different CSA values ($p <0.05$) except for the T2w/MT-on pair. 

Despite significant paired differences across contrasts for both \texttt{soft\_all} and \textblue{\texttt{hard\_all\_SoftSeg}} models, the variability of CSA across contrasts did indeed reduce significantly, as described in the next section.

\subsection{Comparison with baselines}
\label{subsec:baselines-comp}

\begin{figure}[htbp!]
\centering
\includegraphics[width=.485\textwidth]{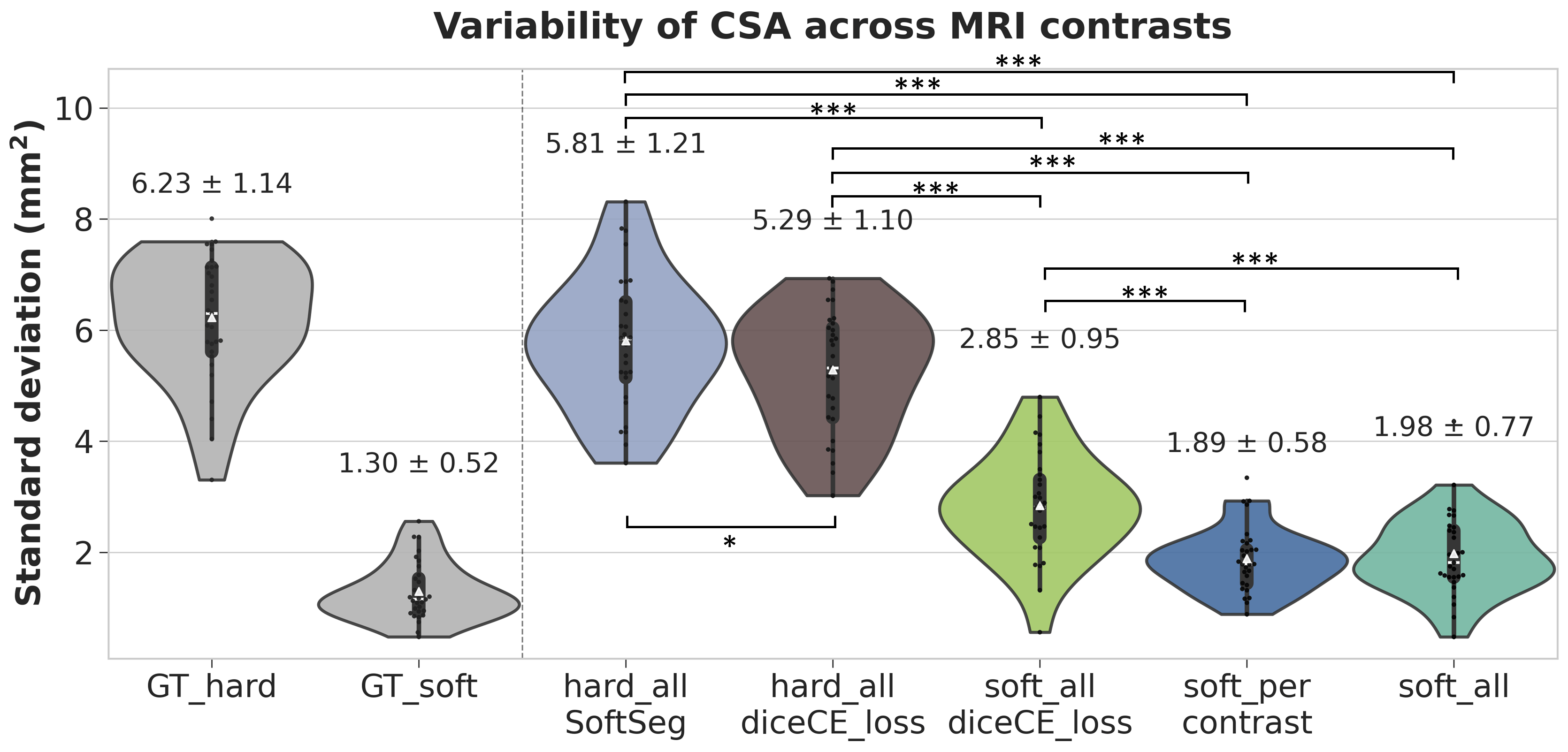}
\caption{Standard deviation of CSA averaged across C2-C3 vertebral levels compared to the baselines (the lower the better). \textblue{\texttt{hard\_all\_SoftSeg} refers to the single model trained using all contrasts with hard GT and the SoftSeg training approach \citep{Gros2021-ms}}, \textblue{\texttt{hard\_all\_diceCE\_loss} refers to the single model trained with the DiceCE loss and hard individual GT}, \textblue{\texttt{soft\_all\_diceCE\_loss} refers to the single model trained with the Dice CE loss and }soft GT, \texttt{soft\_per\_contrast} refers to the mean of 6 individual models trained on 6 contrasts with soft GT, and \texttt{soft\_all} refers to the single model trained using all contrasts with soft GT. White triangle marker shows the mean. * $p <0.05$, ** $p < 0.01$, *** $p < 0.001$ (two-sided Bonferroni-corrected non-parametric Wilcoxon signed-rank test).}
\label{fig:baselines-std-csa-comparison}
\end{figure}

Figure \ref{fig:baselines-std-csa-comparison} compares the performance of the \texttt{soft\_all} model with the baselines in terms of the STD of CSA across contrasts. The STD is computed across all 6 contrasts for each participant and each test participant is represented by an individual point in the scatter plot. Starting with the GT violin plots on the left (gray), we observe that the root cause of CSA variability can be mitigated using soft average segmentations as the GT during training. This is also supported by the fact that the \textblue{\texttt{hard\_all\_SoftSeg}} model, trained on binary GT, results in higher STD when compared to its soft counterparts. Within the models trained using soft GT, the performance of a single model trained on all contrasts (\texttt{soft\_all}) is similar to the average of 6 models trained individually on each contrast (\texttt{soft\_per\_contrast}). Lastly, on comparing the effect of training with \textblue{DiceCE loss (\texttt{soft\_all\_diceCE\_loss})} and adaptive wing loss (\texttt{soft\_all}), we observe \textblue{significantly} lower CSA STD across contrasts when using the regression-based adaptive wing loss \textblue{(p $<$ 0.001)}. 

\begin{figure}[htbp!]
\centering
\includegraphics[width=.485\textwidth]{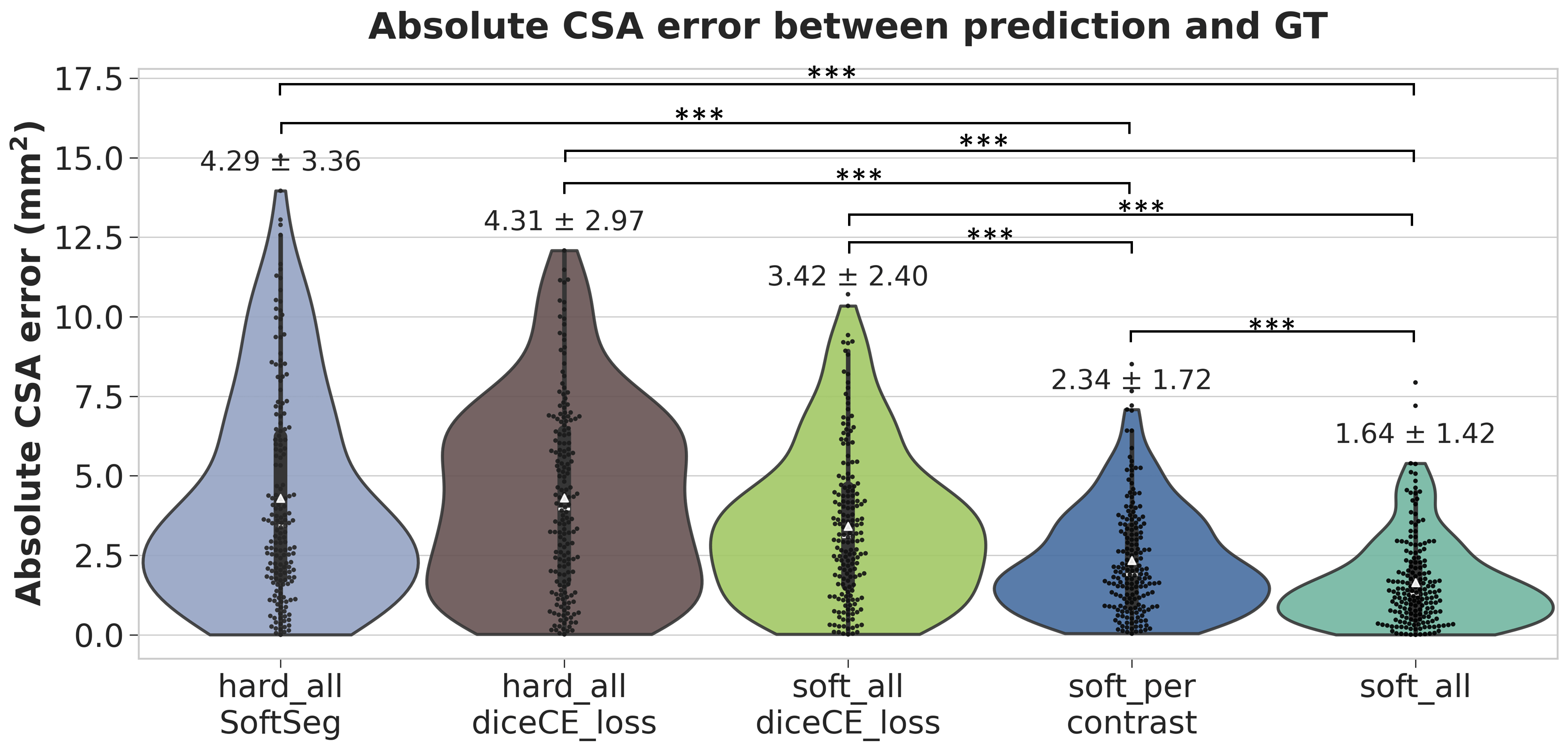}
\caption{Mean absolute CSA error compared against the baselines. \textblue{\texttt{hard\_all\_SoftSeg} refers to the single model trained using all contrasts with hard GT and the SoftSeg training approach \citep{Gros2021-ms}}, \textblue{\texttt{hard\_all\_diceCE\_loss} refers to the single model trained with the Dice CE loss and hard individual GT}, \textblue{\texttt{soft\_all\_diceCE\_loss} refers to the single model trained with the DiceCE loss and }soft GT, \texttt{soft\_per\_contrast} refers to the mean of 6 individual models trained on 6 contrasts with soft GT, and \texttt{soft\_all} refers to the single model trained using all contrasts with soft GT. White triangle marker shows the mean. ** $p < 0.01$, *** $p < 0.001$ (two-sided Bonferroni-corrected non-parametric pairwise Wilcoxon signed-rank test between \texttt{soft\_all} and the \textblue{4} other methods.}
\label{fig:baselines-abs-csa-error-comparison}
\end{figure}

While the previous figure showed the STD of CSA across contrasts for each of the test participant, Figure \ref{fig:baselines-abs-csa-error-comparison} compares the absolute CSA error between the prediction and the GT for our model and the baselines. The points in the scatter plots represent each test image and the mean CSA error is given on top of the violin plots. The superior performance of \texttt{soft\_all} suggests that a combination of soft segmentations \textblue{GT} along with adaptive wing loss is crucial for mitigating CSA variability. 
When comparing the CSA errors between \texttt{soft\_per\_contrast} and \texttt{soft\_all} models, we observe significantly lower \textblue{CSA} errors (p $<$ 0.001) with the latter as also depicted in the violin plot containing a high density of scatter points between $0 - 2$ $\text{mm}^2$ range.
A few more plots comparing the variability in absolute CSA errors across each contrast between the baselines are shown in Section 1.2 of the supplementary material.

\subsection{Comparison with the state of the art}
\label{subsec:sota-comp}

Given the results from the comparison with baselines in the previous section, we considered \texttt{soft\_all} as the best model for subsequent comparisons with the existing SOTA methods. Recall that \texttt{soft\_all} denotes the model that has been trained with soft GT averaged from the segmentations of each of the 6 contrasts \textit{and} adaptive wing loss was used as the loss function.  Table \ref{tab:sota-comparison-metrics} shows the quantitative results for the proposed segmentation model (\texttt{soft\_all}), and the existing SOTA methods. 
The \texttt{soft\_all} model outperforms the SOTA methods in terms on Dice coefficient \textblue{and performs slightly worse in terms of RVE compared to DeepSeg2D and average surface distance (ASD) compared to nnUNet2D, respectively. 
DeepSeg2D and both nnUNet models show under-segmentation (reflected by the negative RVE value) while PropSeg, DeepSeg3D, BigAug and SoftSeg models present over-segmentation.} Also note the relatively low Dice STD for the \texttt{soft\_all} model, suggesting higher robustness (i.e., other models fail more often in presence of difficult images).

\begin{table}[htbp!]
    \centering
    \caption{Quantitative comparison of spinal cord segmentations for the state of the art methods on the test set (294 images) averaged across all contrasts. RVE stands for Relative Volume Error and ASD stands for Average Surface Distance. }
    \vspace{7pt}
    \resizebox{.5\textwidth}{!}{
    \begin{tabular}{l c c c}
        \toprule
        Methods & Dice ($\uparrow$) & RVE \%  & ASD ($\downarrow$) \\
        \cmidrule{2-4}
                  & Opt. value: 1     & Opt. value: 0         & Opt. value: 0 \\
        \midrule
        PropSeg \citep{De_Leener2014-tk}    &  $0.85 \pm 0.15$  & $7.18 \pm 32.59$   & $0.49 \pm 3.92$ \\
        DeepSeg3D \citep{Gros2019-uq}      &  $0.85 \pm 0.13$ &  $18.25 \pm 49.12$ & $0.12 \pm 0.31$\\
        \textblue{\texttt{hard\_all\_BigAug} \citep{Ling2020GDL}}      &  \textblue{$0.92 \pm 0.02$} &  \textblue{$3.16 \pm 6.07$} & \textblue{$0.02 \pm 0.10$}\\
        DeepSeg2D \citep{Gros2019-uq}      &  $0.95 \pm 0.03$ &  $\textbf{-0.24} \pm \textbf{8.64}$  & $0.06 \pm 0.31$\\
        nnUNet3D \citep{Isensee2021-ja}       & $0.95 \pm 0.02$   & $-2.11 \pm 4.43$   & $0.04 \pm 0.29$ \\
        \textblue{nnUNet2D \citep{Isensee2021-ja}}       & \textblue{$0.95 \pm 0.02$}   & \textblue{$-1.85 \pm 4.46$}   & \textblue{$0.02 \pm 0.10$} \\
        \textblue{\texttt{hard\_all\_SoftSeg} \citep{Gros2021-ms}}              & \textblue{$0.96 \pm 0.02$}   & \textblue{$1.77 \pm 5.74$}   & \textblue{$\textbf{0.01} \pm \textbf{0.05}$} \\
        \texttt{soft\_all} (ours)                    &  $ \textbf{0.96} \pm \textbf{0.01}$   & $0.6 \pm 3.82$   & $0.03 \pm 0.12$ \\
        \bottomrule
    \end{tabular}
    }
    \label{tab:sota-comparison-metrics}
\end{table}

\begin{figure}[t]
\centering
\includegraphics[width=.485\textwidth]{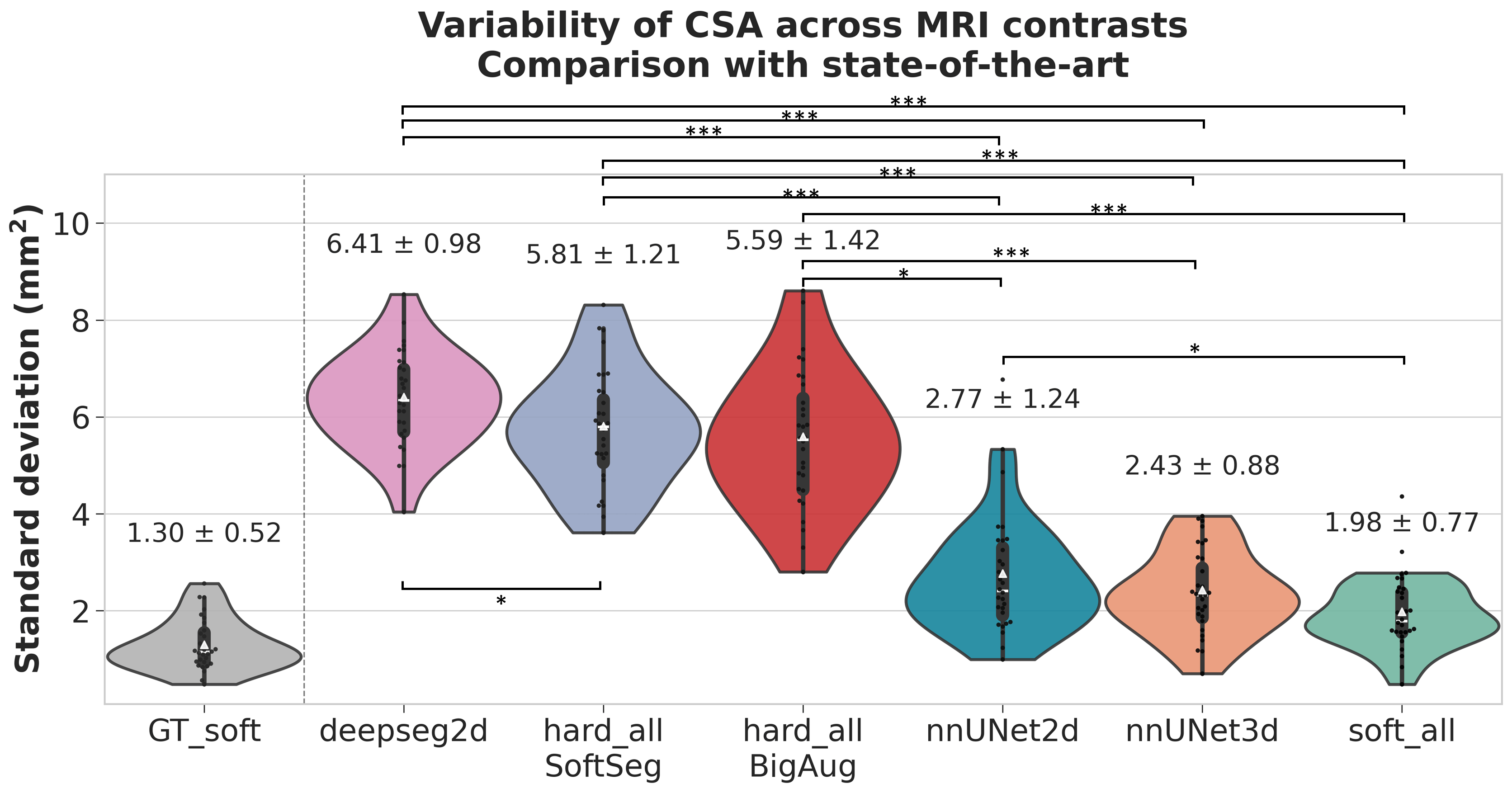}
\caption{Standard deviation of CSA between C2-C3 vertebral levels for \textblue{DeepSeg2D, \texttt{hard\_all\_SoftSeg},  \texttt{hard\_all\_BigAug}, nnUNet 2D/3D, and our model \texttt{soft\_all}}.   * $p < 0.05$, ** $p < 0.01$, *** $p < 0.001$ (two-sided Bonferroni-corrected non-parametric pairwise Wilcoxon signed-rank test between  \textblue{each pair of methods.}}
\label{fig:sota-std-csa-comparison}
\end{figure}

Figure \ref{fig:sota-std-csa-comparison} compares the STD of CSA across contrasts for our best model and the existing methods. \textblue{
DeepSeg2D has the highest STD across contrasts, followed by \texttt{hard\_all\_SoftSeg} and \texttt{hard\_all\_BigAug}.
}
Interestingly, despite the hard requirement of having binarized GT and training with DiceCE loss, \textblue{both nnUNet models achieved similar STD across contrasts with the 3D model showing lower STD across contrasts. Overall, both nnUNet 2D and 3D models showed higher STD when compared to our best model \texttt{soft\_all}}.
\textblue{In Supplementary Figure S8, we show that for nnUNet3D}, the relatively lower STD values do not necessarily correspond to lower absolute CSA errors across individual contrasts (as our model achieves in Figure \ref{fig:soft-all-abs-csa-error-per-contrast}). 


\begin{figure}[htbp!]
\centering
\includegraphics[width=.485\textwidth]{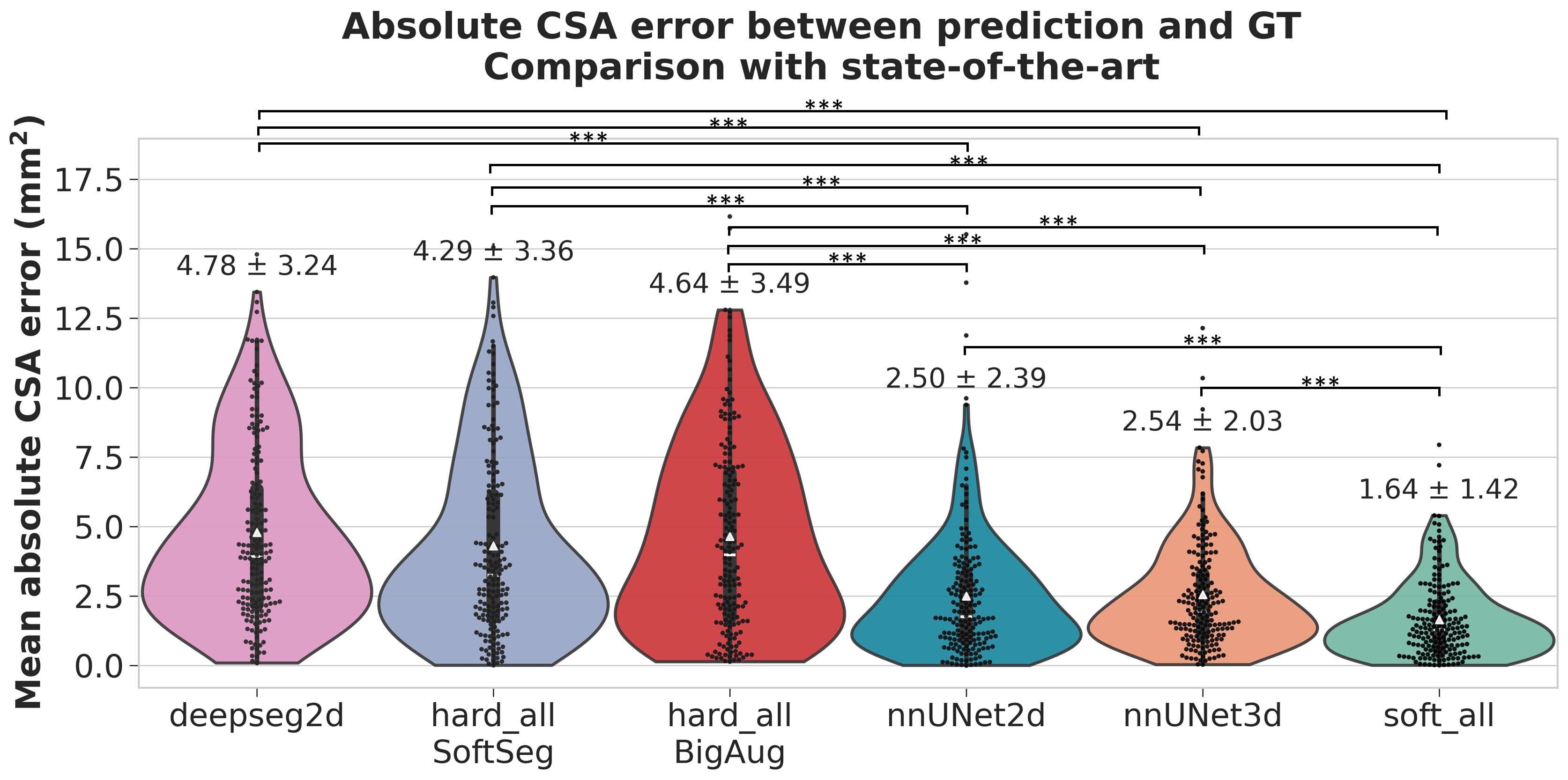}
\caption{\textblue{Mean absolute CSA error for DeepSeg 2D,  \texttt{hard\_all\_SoftSeg}, \texttt{hard\_all\_BigAug}, nnUNet 2D/3D, and our model \texttt{soft\_all}. * $p < 0.05$, ** $p < 0.01$, *** $p < 0.001$ (two-sided Bonferroni-corrected non-parametric pairwise Wilcoxon signed-rank test between each pair of methods.}}
\label{fig:sota-abs-csa-error-comparison}
\end{figure}

Figure \ref{fig:sota-abs-csa-error-comparison} shows a comparison between our best model (\texttt{soft\_all}) and the other methods in terms of the absolute CSA error. Similar to the trend observed in Figure \ref{fig:sota-std-csa-comparison}, we noted that \texttt{soft\_all} achieves the lowest mean absolute CSA error with $1.64 \pm 1.42$ $\text{mm}^2$ across all test images. 
A closer look at the absolute CSA errors per contrast for each method is shown in Section 1.3 of the supplementary material. 
\textblue{In supplementary Figures S9 and S10, we show the comparison with all the methods including DeepSeg 3D and PropSeg to ovoid overcrowding Figures \ref{fig:sota-std-csa-comparison} and \ref{fig:sota-abs-csa-error-comparison}.}


\begin{figure}[htbp!]
\centering
\includegraphics[width=.485\textwidth]{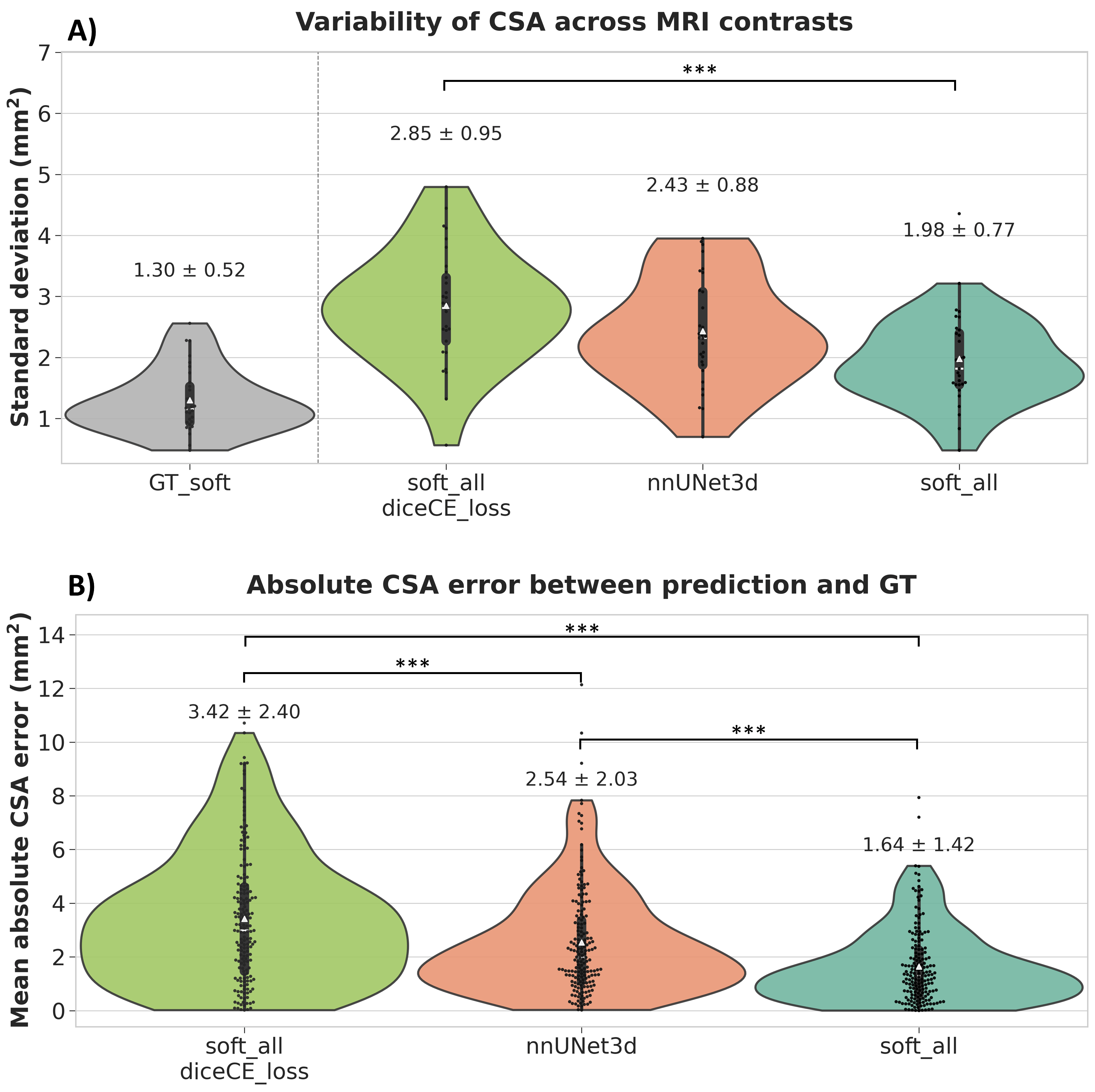}
\caption{\textblue{Comparison of CSA estimation between models trained on soft masks. 
\textbf{A)} Standard deviation of CSA between C2-C3 vertebral levels for \texttt{soft\_all\_diceCE\_loss}, nnUNet 3D, and our model \texttt{soft\_all}.   * $p < 0.05$, ** $p < 0.01$, *** $p < 0.001$ (two-sided Bonferroni-corrected non-parametric pairwise Wilcoxon signed-rank test between \texttt{soft\_all}, nnUNet 3D and \texttt{soft\_all\_diceCE\_loss}). 
\textbf{B)} Mean absolute CSA error for \texttt{soft\_all\_diceCE\_loss}, nnUNet 3D, and our model \texttt{soft\_all}. * $p < 0.05$, ** $p < 0.01$, *** $p < 0.001$ (two-sided Bonferroni-corrected non-parametric pairwise Wilcoxon signed-rank test between \texttt{soft\_all}, nnUNet 3D and \texttt{soft\_all\_diceCE\_loss}.}}
\label{fig:soft-models-comparison}
\end{figure}

\textblue{
To understand the impact of data augmentation and loss functions on training with soft masks, we compared our models trained using DiceCE and adaptive wing loss with nnUNet in Figure \ref{fig:soft-models-comparison}.
Keeping the loss function fixed (i.e. DiceCE), we can observe that data augmentation transforms in nnUNet3D play a stronger role as it achieves lower STD and absolute CSA errors compared to the \texttt{soft\_all\_diceCE\_loss} model. 
However, as shown by \texttt{soft\_all}, switching to the adaptive wing loss irrespective of data augmentation transforms leads to further reduction in the CSA variability across contrasts. 
}

\begin{figure}[htbp!]
\centering
\includegraphics[width=.475\textwidth]{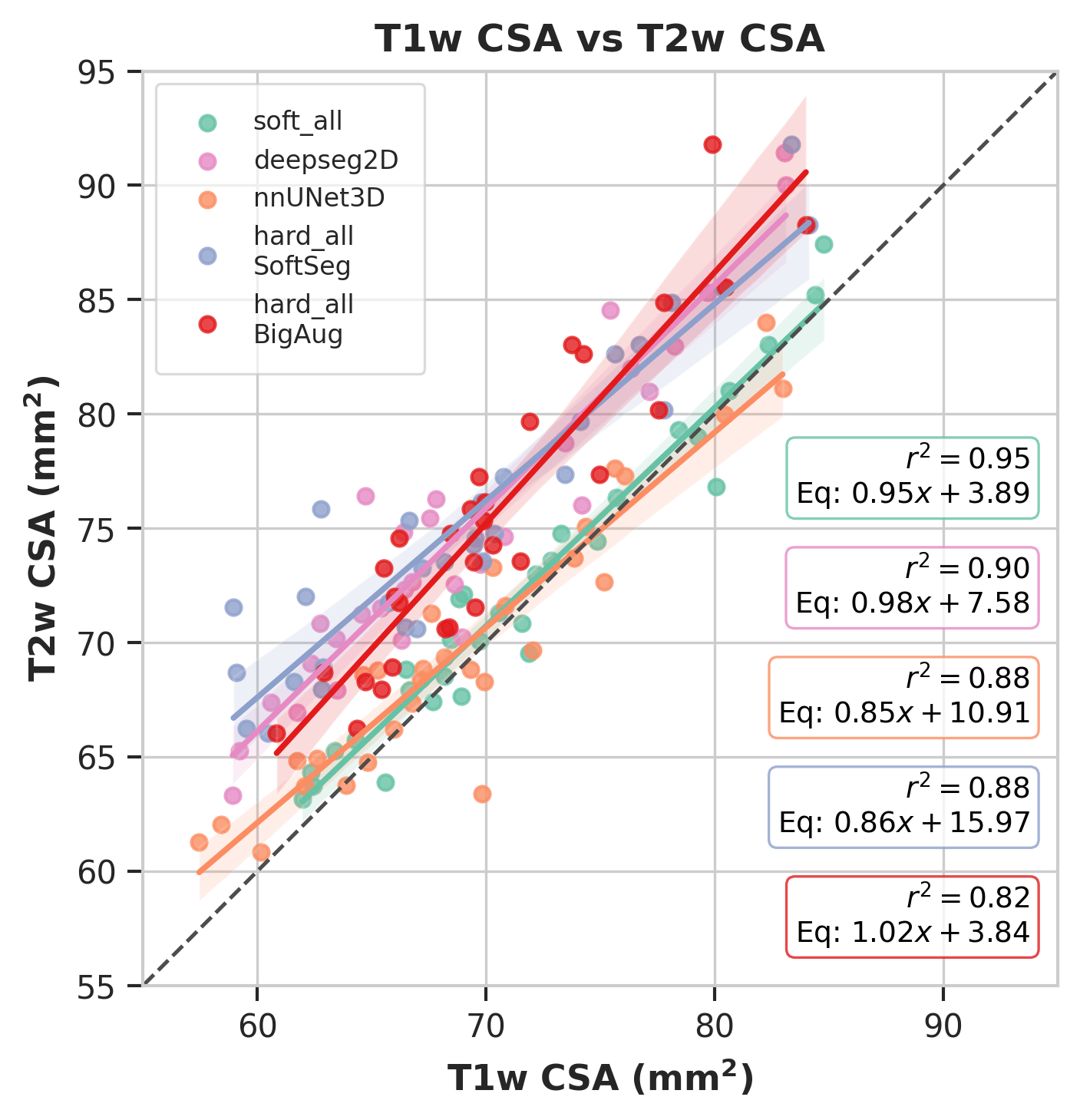}
\caption{\textblue{Level of agreement between T1w and T2w CSA for the best-performing SOTA methods}. Each point represents one participant. The black dashed line represents perfect agreement between the CSA of T1w and T2w contrasts.}
\label{fig:pairplot-t1-t2}
\end{figure}

\textblue{
In Figure \ref{fig:pairplot-t1-t2}, we plotted the level of agreement between the CSA estimated by our model (\texttt{soft\_all}) and the SOTA methods on T1w and T2w contrasts. The models trained with individual hard masks (namely, \texttt{deepseg2D}, \texttt{hard\_all\_SoftSeg}, and \texttt{hard\_all\_BigAug}) showed large discrepancies between the estimated CSA. Interestingly, the models trained with soft masks obtained by averaging all contrasts moved closer to the diagonal line representing perfect alignment between T1w/T2w CSA. Within \texttt{soft\_all} and \texttt{nnUNet3D}, our model achieves better alignment between the two contrasts, thus confirming that it 
reduces discrepancies between these two popular contrasts in spinal cord imaging. 
The correlation plots for the remaining pairs of contrasts for \texttt{soft\_all} are shown in Figure S1 in the supplementary material.
}

\subsection{Generalization to unseen data}
\label{subsec:generalization}
\begin{figure}[htbp!]
\centering
\includegraphics[width=.485\textwidth]{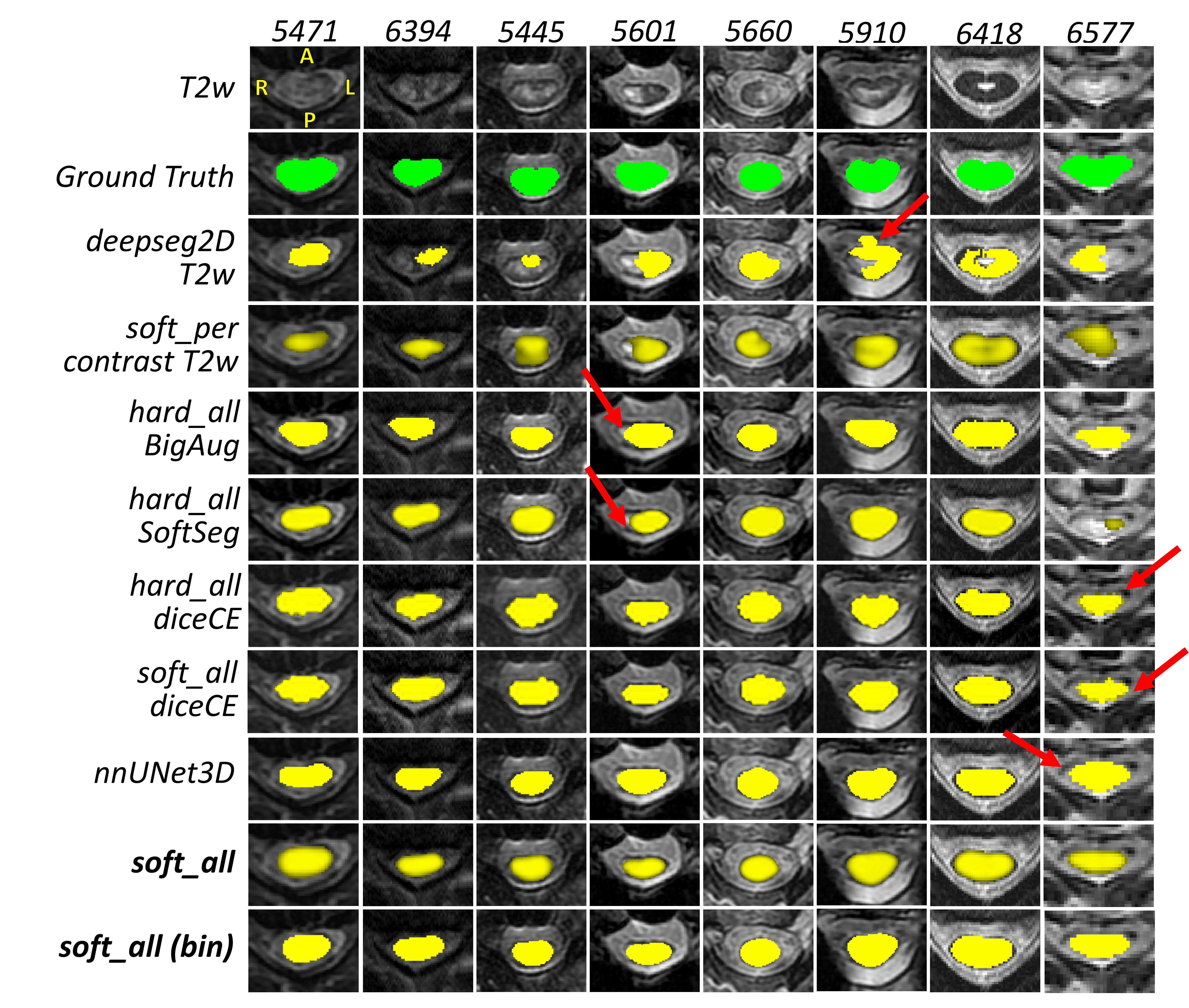}
\caption{T2w axial slices with the overlaid GT (green) and model predictions (yellow) in 8 patients with  traumatic spinal cord injury (\texttt{sci-t2w} dataset). The red arrow indicates \textblue{example of segmentation errors}. Soft segmentations are clipped at 0.5. \texttt{soft\_all(bin)} represents the \texttt{soft\_all} binarized at 0.5 for better visual comparison with the GT and hard segmentation methods.}
\label{fig:sci-comparison}
\end{figure}

Figure \ref{fig:sci-comparison} shows the predictions for 8 representative patients with spinal cord injury from the \texttt{sci-t2w} dataset. Despite the presence of spinal cord lesions, we notice that \texttt{soft\_all} and nnUNet models were able to correctly segment the spinal cord, while DeepSeg2D T2w and \texttt{soft\_per\_contrast} T2w models under-segmented the spinal cord (except one over-segmentation pointed by the red arrow).  
\textblue{The \texttt{hard\_all\_SoftSeg},  \texttt{hard\_all\_BigAug},  \texttt{hard\_all\_diceCE} and \texttt{soft\_all\_diceCE} models showed under-segmentation of the hyperintense lesions in the spinal cord as indicated by the red arrows.
}

\begin{figure}[htbp!]
\centering
\includegraphics[width=.485\textwidth]{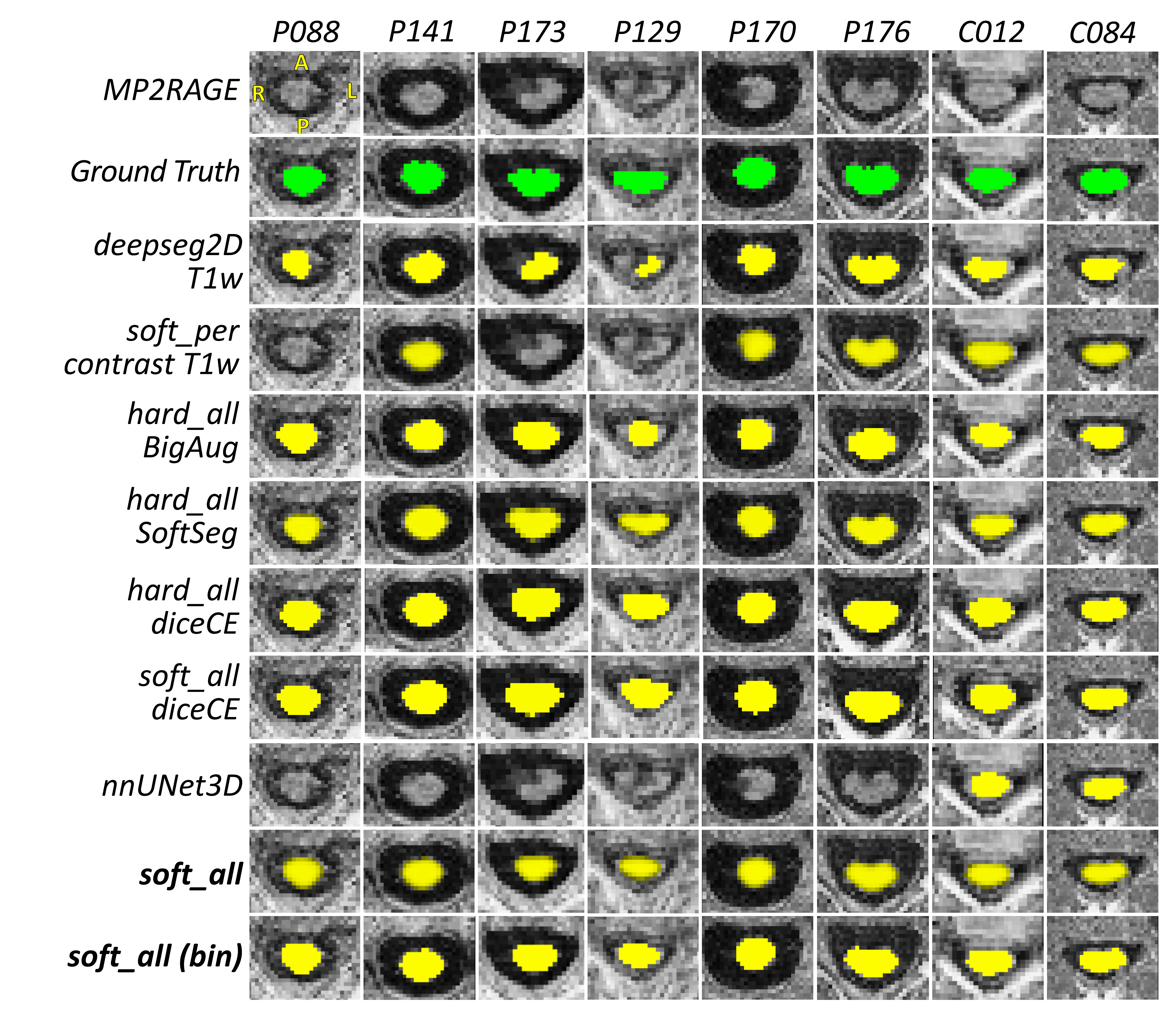}
\caption{MP2RAGE axial slices with the overlaid GT (green) and model predictions (yellow) in 6 patients (P) with multiple sclerosis lesions and 2 healthy controls (C) (\texttt{ms-mp2rage} dataset). Soft segmentations are clipped at 0.5. \texttt{soft\_all(bin)} represents the \texttt{soft\_all} binarized at 0.5.}
\label{fig:mp2rage-comparison}
\end{figure}

Figure \ref{fig:mp2rage-comparison} shows the predictions for 6 representative patients with multiple sclerosis and 2 healthy participants from the \texttt{ms-mp2rage} dataset. Despite the presence of spinal cord MS lesions, we observe that the \texttt{soft\_all} model was able to correctly segment the spinal cord, while DeepSeg2D T1w model under-segmented the spinal cord typically at the lesion location. The \texttt{soft\_per\_contrast} T1w and nnUNet models were unable to properly segment the spinal cord in the presence of hypointense lesions. \textblue{ \texttt{hard\_all\_BigAug} showed some under-segmentation of the location of the hypointense lesions. \texttt{hard\_all\_SoftSeg},   \texttt{hard\_all\_diceCE} and  \texttt{soft\_all\_diceCE} showed similar performance and were able to properly segment the spinal cord. }

\begin{figure}[htbp!]
\centering
\includegraphics[width=.485\textwidth]{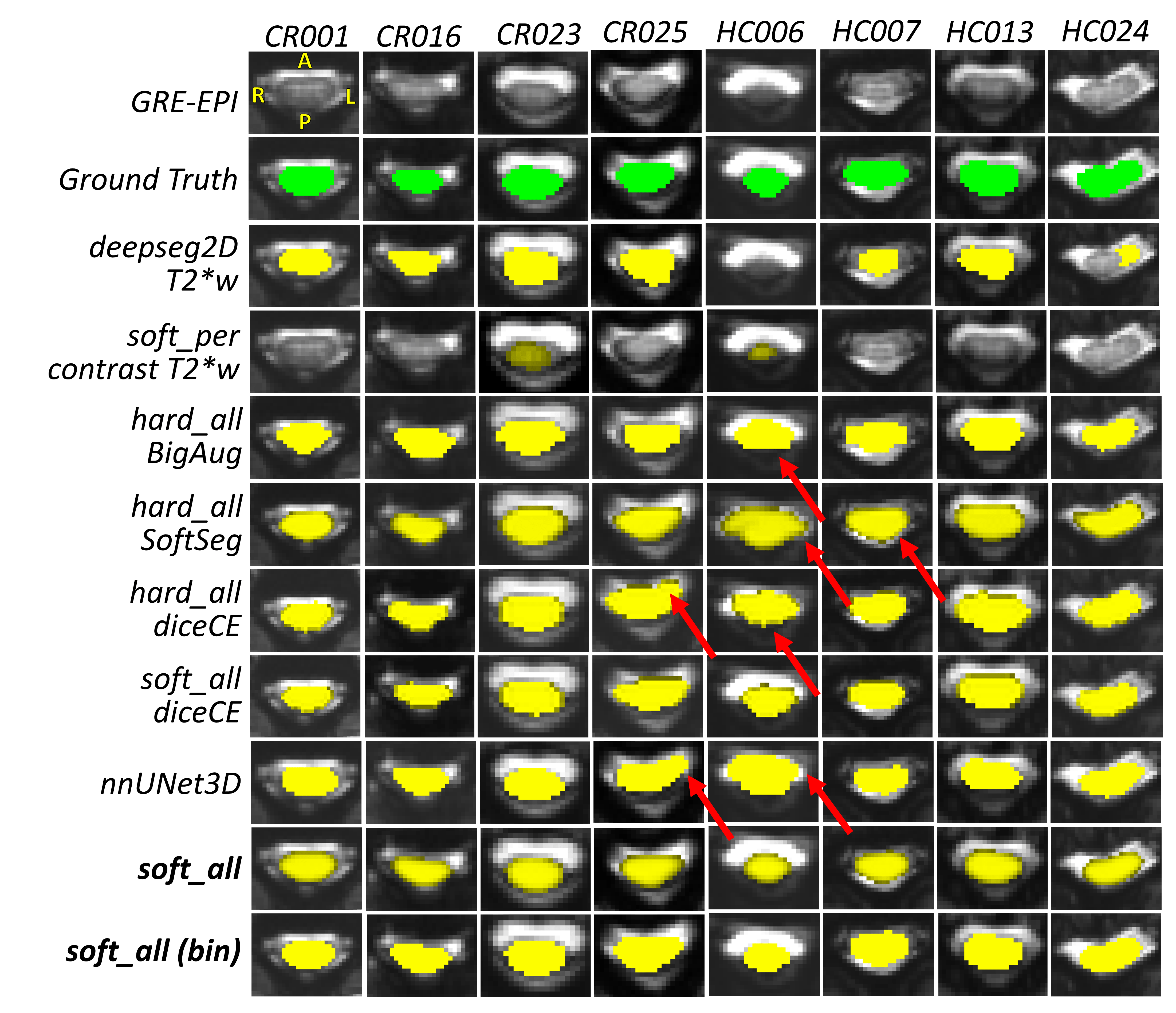}
\caption{GRE-EPI axial slices with the overlaid GT (green) and model predictions (yellow) of spinal cord segmentation of 4 patients with cervical radiculopathy (CR) and 4 healthy controls (HC). Soft segmentations are clipped at 0.5. \texttt{soft\_all(bin)} represents the \texttt{soft\_all} binarized at 0.5. Red arrows indicate \textblue{examples of segmentation errors}.}
\label{fig:gre-epi-comparison}
\end{figure}

Figure \ref{fig:gre-epi-comparison} shows the predictions for 4 representative patients with cervical radiculopathy and 4 healthy controls from the \texttt{radiculopathy-epi} dataset. The \texttt{soft\_all} model was able to correctly segment the spinal cord even with the poor image quality of the gradient echo EPI. In contrast, the \texttt{soft\_per\_contrast} T2*w was unable to segment the spinal cord in almost all cases. The DeepSeg2D T2*w and nnUNet performed slightly worst than \texttt{soft\_all}, notably in slices affected by signal drop out (e.g., HC013), and nnUNet, \textblue{\texttt{hard\_all\_diceCE}, \texttt{hard\_all\_SoftSeg} }had a tendency to leak into the cerebrospinal fluid (red arrows).
\textblue{The \texttt{hard\_all\_BigAug} model had trouble with getting the shape of the spinal cord, mainly in the presence of signal drop-out, shown by the red arrow.}
\textblue{
Representative examples of generalization to the FLAIR contrast can be found in the supplementary material Figure S11.}

Table \ref{tab:other-method-dice} presents the quantitative metrics for the models' performance on all three \textit{external} datasets. We do not report the Dice coefficients for DeepSeg2D since the GT masks were generated using this model with manual corrections on the slices presenting over/under-segmentations, hence biasing the scores. 
We observed that the nnUNet3D performed well on the T2w and GRE-EPI contrasts while performing poorly on the MP2RAGE "UNI" contrast (Dice = 0.24 $\pm$ 0.25) due to large under-segmentation overall (RVE = $-82.51 \pm 19.27 \%$) \textblue{while both models using DiceCE (\texttt{hard\_all\_DiceCE} and \texttt{soft\_all\_DiceCE}) performed well on MP2RAGE "UNI" contrast, as we can also observe in Figure ~\ref{fig:mp2rage-comparison}}.
\textblue{Both BigAug \citep{Ling2020GDL} and SoftSeg \citep{Gros2021-ms} models outperformed nnUNet3D only for the MP2RAGE contrast.}
The \texttt{soft\_per\_contrast} model performed poorly on GRE-EPI (Dice = 0.29 $\pm$ 0.18) but perfomed well on the MP2RAGE "UNI" and T2w data. In terms of RVE, we see that nnUNet 3D consistently shows under-segmentation on all datasets. 
The \texttt{soft\_all} model outperforms all the tested methods for \texttt{radiculopathy-epi} datasets in terms of Dice coefficients and performs similar to nnUNet on the \texttt{sci-t2w} dataset, \textblue{and similar to SoftSeg on the \texttt{ms-mp2rage} dataset}.

\setlength{\tabcolsep}{8pt}
\begin{table}[htbp!]
    \centering
    \caption{Comparison of quantitative metrics between SOTA methods for spinal cord segmentation on unseen datasets. $n$ refers to the number of participants.}
    \vspace{7pt}
    \resizebox{.5\textwidth}{!}{
    \begin{tabular}{l c c c}
        \toprule
           &   \multicolumn{3}{c}{Dataset \texttt{sci-t2w} ($n = 80$)} \\
        \midrule
        Methods  & Dice ($\uparrow$) & RVE \%  & ASD ($\downarrow$) \\    
        \cmidrule{2-4}
                  & Opt. value: 1     & Opt. value: 0         & Opt. value: 0 \\
        \midrule
        \textblue{\texttt{hard\_all\_DiceCE}}               
        & \textblue{$0.86 \pm 0.06$}     & \textblue{$-8.83 \pm 11.95$}      & \textblue{$0.03 \pm 0.12$}  \\
        \textblue{\texttt{hard\_all\_BigAug}}               
        & \textblue{$0.83 \pm 0.08$}     & \textblue{$-\textbf{5.67} \pm \textbf{14.84}$}      & \textblue{$0.03 \pm 0.16$}  \\
        \textblue{\texttt{hard\_all\_SoftSeg}}               
        & \textblue{$0.85 \pm 0.09$}     & \textblue{$-14.44 \pm 14.25$}      & \textblue{$0.17 \pm 0.98$}  \\
       \texttt{soft\_per\_contrast}             & $0.84 \pm 0.09$   & $-16.45 \pm 11.19$ & $0.04 \pm 0.19$ \\
        \texttt{nnUNet3D}                                  
        & $\textbf{0.87 }\pm \textbf{0.05}$     & $-15.20 \pm 8.59$       & $\textbf{0.01} \pm \textbf{0.03}$ \\
       \textblue{\texttt{soft\_all\_DiceCE}}             & \textblue{$\textbf{0.87} \pm \textbf{0.05}$}   & \textblue{$-8.30 \pm 10.72$} & \textblue{$0.02 \pm 0.10$} \\
       \texttt{soft\_all}                       & $0.86 \pm 0.07$   & $-16.14 \pm 9.80$ & $0.01 \pm 0.06 $\\
        
       \midrule 
          &  \multicolumn{3}{c}{Dataset \texttt{ms-mp2rage} ($n = 283$)} \\
       \midrule  
        \textblue{\texttt{hard\_all\_DiceCE}}               
        & \textblue{$0.92 \pm 0.03$}     & \textblue{$8.51 \pm 6.15$}      & \textblue{$0.01 \pm 0.03$}  \\
        \textblue{\texttt{hard\_all\_BigAug}}               
        & \textblue{$0.89 \pm 0.05$}     & \textblue{$-2.38 \pm 9.69$}      & \textblue{$0.20 \pm 0.75$}  \\
        \textblue{\texttt{hard\_all\_SoftSeg}}               
        & \textblue{$\textbf{0.93} \pm \textbf{0.02}$}     & \textblue{$7.57 \pm 5.61$}      & \textblue{$0.01 \pm 0.08$}  \\
        \texttt{soft\_per\_contrast}    & $0.83 \pm 0.14$     &  $-14.38 \pm 22.44$     & $0.15 \pm 0.44$   \\
        \texttt{nnUNet3D}   
        & $0.24 \pm 0.25$     & $-82.51 \pm 19.27$      & $5.09 \pm 18.15 $  \\
       \textblue{\texttt{soft\_all\_DiceCE}}             & \textblue{$0.90 \pm 0.04$}   & \textblue{$17.56 \pm 7.54$} & \textblue{$0.01 \pm 0.04$} \\
        \texttt{soft\_all}              & $0.93 \pm 0.03$    &  $\textbf{6.88} \pm \textbf{6.09}$     & $\textbf{0.01} \pm \textbf{0.03}$  \\

        \midrule 
          &  \multicolumn{3}{c}{Dataset \texttt{radiculopathy-epi} ($n = 52$)} \\
       \midrule     
        \textblue{\texttt{hard\_all\_DiceCE}}               
        & \textblue{$0.87 \pm 0.04$}     & \textblue{$20.98 \pm 11.56$}      & \textblue{$0.01 \pm 0.02$}  \\
        \textblue{\texttt{hard\_all\_BigAug}}               
        & \textblue{$0.87 \pm 0.03$}     & \textblue{$-2.52 \pm 10.4$}      & \textblue{$0.01 \pm 0.01$}  \\
        \textblue{\texttt{hard\_all\_SoftSeg}}               
        & \textblue{$0.88 \pm 0.04$}     & \textblue{$17.14 \pm 10.85$}      & \textblue{$0.02 \pm 0.02$}  \\
        \texttt{soft\_per\_contrast}    & $0.29 \pm 0.18$     & $-80.64 \pm 12.82$    & $0.58 \pm 1.06$   \\
        \texttt{nnUNet3D}                          & $0.90 \pm 0.04$     & $-4.66 \pm 8.79$      & $0.01 \pm 0.02$   \\
       \textblue{\texttt{soft\_all\_DiceCE}}             & \textblue{$0.88 \pm 0.04$}   & \textblue{$10.53 \pm 14.36$} & \textblue{$0.01 \pm 0.01$} \\
        \texttt{soft\_all}              & $\textbf{0.90} \pm \textbf{0.03}$     & $\textbf{2.83} \pm \textbf{11.22}$      & $\textbf{0.01} \pm \textbf{0.01}$   \\
       
        \bottomrule
    \end{tabular}
    }
    \label{tab:other-method-dice}
\end{table}

\subsection{Inference time}
\label{subsec:inference-time}
\begin{figure}[htbp!]
\centering
\includegraphics[width=.375\textwidth]{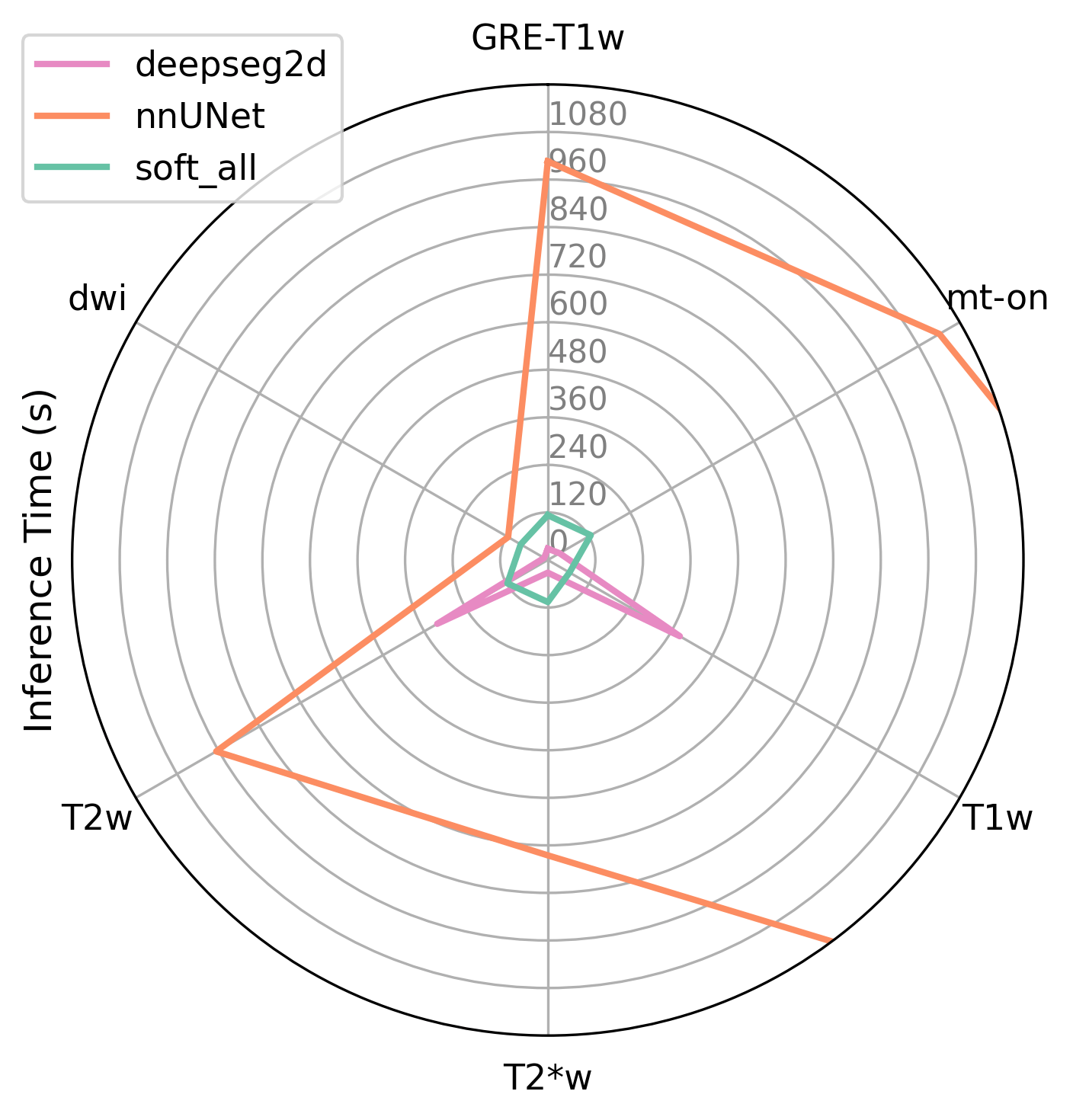}
\caption{Inference times (in seconds) averaged across test participants for DeepSeg2D, nnUNet\textblue{3D}, and \texttt{soft\_all} for all contrasts.}
\label{fig:inference-times-comparison}
\end{figure}

Figure \ref{fig:inference-times-comparison} compares the average inference time across participants for the DeepSeg2D, nnUNet\textblue{3D} and \texttt{soft\_all} methods. The inference is done on a CPU (Intel  Xeon E7-4850 @ 2.10GHz) and the time is shown in seconds. \texttt{soft\_all} takes up to 2 minutes per prediction on average irrespective of the contrast, whereas nnUNet \textblue{3D} 's inference times are highly variable and longer. For example, nnUNet\textblue{3D}  takes about 3000 seconds (50 mins) for segmenting a T1w image (not shown in the plot for clarity), while \texttt{soft\_all} requires only about 2 minutes. Unsurprisingly, the inference time largely depends on the size of the input volume. 
In addition to obtaining good segmentations, the average inference time per image is an important factor for consideration before deploying the model in real-world clinical settings. Models such as nnUNet\textblue{3D} requiring long inference times are impractical when used on large cohorts.

\section{Discussion} 
\label{sec:discussion}
We presented an automatic method for the contrast-agnostic soft segmentation of the spinal cord. Starting with the creation of GT masks, we proposed a new framework for generating a unique (soft) GT that represents the segmentations across various MR contrasts. Using these soft GT masks along with the corresponding images of all contrasts as inputs, we trained a U-Net model with aggressive data augmentation, regression-based adaptive wing loss, and NormReLU as the final activation function for spinal cord segmentation. We evaluated our model against 3 categories of baseline models to assess the impact of soft vs. hard GT masks, training a single model with all contrasts vs. training one model per contrast, and the impact of the loss functions on the subsequent morphometric measures (i.e., CSA). We compared our model's performance with the SOTA methods for spinal cord segmentation, namely, PropSeg, DeepSeg, \textblue{SoftSeg, BigAug} and nnUNet. To demonstrate our model's domain generalization capabilities on unseen contrasts and images with pathology (lesions), we presented a qualitative comparison against spinal cord segmentations of the SOTA methods. Lastly, we provided a graphical illustration of the average inference times of the existing methods and highlighted that our model takes only a fraction of the time per image irrespective of the contrast, while obtaining better segmentations that reduce the bias/variability in morphometric measures.

\subsection{\textblue{Preprocessing for Soft GT} }
\textblue{
Our preprocessing framework to generates a unique soft segmentation from individual hard segmentations. The procedure required aligning all the images to the T2w image space, including resampling and reorientation. The soft segmentation was then brought back in each contrast's native space. Then, in order to have a fixed patch size for training with all contrasts, we performed an up-sampling to 1 mm isotropic on all images and labels as a preprocessing step during data augmentation. One could question the reason for this up-down-up sampling in the preprocessing workflow, a few points are discussed below:
}

\vspace{-10pt}
\textblue{
\begin{itemize}
    \item \textbf{Comparison with baselines}: In order to provide a fair comparison with the baselines, we ensured that all methods used the ground truth defined in the native space of each contrast.
    \item \textbf{Computation of CSA and disc labeling}: Continuing the previous point and considering the evaluation of CSA variability across contrasts, it was important to compute the CSA in the native space, notably because of the highly variable spatial resolution across contrasts, which limited our ability to ensure that vertebral coverage would be the same when computed in the native image vs. in the reference image (T2w 0.8mm isotropic).
    \item \textbf{Patch size}: To ensure a uniform patch size during training for all the 6 contrasts, resampling the images to a unique, common, resolution was necessary. After experimenting with various spatial resolutions (trade-off between computational resource and precision), we chose 1 mm as the target resolution.
    \item \textbf{Introduce realistic variability}: One of the advantages of training the model in the native space is that, for a given subject, the spinal cords are not perfectly aligned across contrasts. 
\end{itemize}
}

\subsection{Variability of CSA across contrasts}
 
When comparing the performance on all 6 MR contrasts (Figure \ref{fig:soft-all-abs-csa-error-per-contrast}), DWI and MT-on contrasts yielded the highest absolute CSA error with $2.37  \pm 1.09 $ $ \text{mm}^2$ and $2.38 \pm 1.77 $ $ \text{mm}^2$, respectively. This can be attributed to the fact that DWI and MT-on contrasts have less well defined boundaries between the spinal cord and the cerebrospinal fluid due to the coarser resolution of the images ($0.9 \times 0.9 \times 5$ mm). Furthermore, the presence of susceptibility related artifacts on DWI and MT-on (MT-on is based on GRE readout and hence suffers from signal dropout), and/or the ghosting and motion artifacts that particularly affect MT-on data ~\cite{Cohen-Adad2021-rp} could explain the larger errors.

The agreement between T1w and T2w CSA (Figure \ref{fig:pairplot-t1-t2}) leads to a linear equation given by $0.95x + 3.89$ with an $r^2 = 0.95$, which is very close to the (dashed) identity line. In \cite{Cohen-Adad2021-rp} where \texttt{sct\_deeepseg\_sc} was used for generating spinal cord segmentations, the authors reported poor agreement between CSA computed on T1w and T2w images, regardless of the MR vendor.


Furthermore, training an individual model for each contrast (\texttt{soft\_per\_contrast}) yielded similar performance (albeit with a slightly higher error for each contrast) compared to training a single model for all contrasts (\texttt{soft\_all}), as seen in Figure \ref{fig:baselines-std-csa-comparison}. 
Training an individual model for each contrast does not help in mitigating the CSA variability across contrasts as each model is optimizing for spinal cord segmentations asynchronously, thus leading to different CSA for a given participant despite using soft GT segmentations. 
On the other hand, a single model trained on all contrasts together is exposed to the wide heterogeneity in the images, thus leading to better estimation of the CSA and better generalization (as shown by our results on \texttt{ms-mp2rage} and \texttt{radiculopathy-epi} datasets in Figures \ref{fig:mp2rage-comparison} and \ref{fig:gre-epi-comparison}).
Moreover, from a deployment standpoint, packaging and distributing a single model is more convenient and intuitive for researchers to use. 

Overall, the proposed contrast-agnostic \texttt{soft\_all} model 
outperforms the baselines and existing state-of-the-art spinal cord segmentation methods while minimizing contrast-specific CSA biases.

\subsection{Effects of ground truth masks and loss functions}
Results from Figures \ref{fig:soft-vs-hard-comparison}, \ref{fig:baselines-std-csa-comparison}, and \ref{fig:baselines-abs-csa-error-comparison} demonstrated that the types of GT masks and loss functions used play a crucial role in the (downstream) computation of CSA. 
Using the unique soft GT generated by averaging the segmentations across 6 contrasts compared to traditional hard GT leads to lower CSA variability on the predictions. For more details, see Figure \ref{fig:soft-vs-hard-comparison} (qualitative assessment) and Figure \ref{fig:baselines-std-csa-comparison}  (statistical assessment). Notably, the bias inherent to the individual GT for each contrast is propagated \textit{less} when using the unique soft GT averaged from different contrasts. 

For \texttt{soft\_all}, the CSA did not differ significantly between T1w and T2w contrasts, while it was significantly different between the T1w / T2w and the other contrasts (T2*w, GRE-T1w, MT-on, DWI). Interestingly, T1w and T2w images have similar isotropic resolutions (respectively $1 \times 1 \times 1 \textrm{ mm}$ and $0.8 \times 0.8 \times 0.8 \textrm{ mm}$), whereas the other contrasts feature highly anisotropic axial acquisitions ($> 3$ mm slice thickness). It is therefore possible that excessive partial volume effect along the superior-inferior axis created a bias in the CSA estimation. Another possible explanation of the discrepancy between isotropic and anisotropic scans, is the uncertainty in the estimation of the C2-C3 vertebral labels across contrasts. Since it was not possible to directly label the vertebral levels on the anisotropic scans (because discs are poorly visible), we used the disc labels created from the T2w images and applied the warping field to the labels to the target contrast. This likely resulted in slightly higher CSA STD across contrasts.

\textblue{
When the Dice or DiceCE loss functions \citep{Milletari2016-xj,Isensee2021-ja} were used in combination with hard GT masks, our results suggested that using Dice metric in the training objective is not sufficient for achieving accurate segmentations at the spinal cord / cerebrospinal fluid boundary. In fact, the model trained with Dice loss (\texttt{hard\_all\_BigAug}) showed subtle under-segmentations upon quality control as supported by larger absolute CSA errors in Figure \ref{fig:baselines-abs-csa-error-comparison}.}
Instead, using adaptive wing loss that switches to the logarithmic (non-linear) part of the loss when the error between the prediction and the GT are small, helps the model in refining the segmentations at the boundaries of the spinal cord, thus mitigating the CSA bias across contrasts. Similar observations have been reported about the effectiveness of regression-based \citep{Gros2021-ms} and logarithmic \citep{Kaul2021-ph} loss functions. 

The benefits of using soft GT and adaptive wing loss in our model \texttt{soft\_all} can be seen in Figures \ref{fig:sota-std-csa-comparison} and \ref{fig:sota-abs-csa-error-comparison}.  PropSeg, DeepSeg, \textblue{SoftSeg} and \textblue{BigAug} methods, which used hard GT that are inherently biased due to the procedure of their GT generation, resulted in higher STD across contrasts per participant. As nnUNet does not support soft training yet, using soft segmentations averaged from all contrasts (but binarized at 0.5 threshold), still resulted in slightly higher CSA variability. Furthermore, DeepSeg used Dice loss and nnUNet used DiceCE loss during training, thus explaining the larger errors. 
\textblue{Thus, the subtle yet important difference of training on implicitly obtained soft masks via data augmentation vs. applying the augmentation transforms directly on the soft masks generated from multiple contrasts has a significant downstream impact on the reduction of CSA  variability across contrasts.}

\subsection{Generalization to unseen data}
The proposed \texttt{soft\_all} model demonstrates remarkable generalization on the unseen MP2RAGE "UNI" (\texttt{ms-mp2rage}) and resting state GRE-EPI (\texttt{radiculopathy-epi}) contrasts. Despite being trained only on healthy participants, it performed well on patients with MS lesions, traumatic spinal cord injury and cervical radiculopathy. 
We noticed that the tested models performed similarly on a contrast that was included during training (T2w) as reflected by the Dice coefficients of Table \ref{tab:other-method-dice} and segmentations in Figure \ref{fig:sci-comparison}, even in the presence of traumatic spinal cord injury. 
Surprisingly, there is a marked difference between the qualitative segmentations of (\texttt{soft\_all}) and nnUNet\textblue{3D} (Figures \ref{fig:mp2rage-comparison} and \ref{fig:gre-epi-comparison}) on the \texttt{ms-mp2rage} and \texttt{radiculopathy-epi} datasets, respectively. With the former dataset, nnUNet\textblue{3D} performs poorly by completely missing the spinal cord in the presence of multiple sclerosis lesions, whereas for the latter, we observed cases with over-segmentation of the cord leaking into the cerebrospinal fluid. For both these contrasts, \texttt{soft\_all} obtains accurate segmentations of the spinal cord under the presence of lesions and along the spinal cord boundaries. 

The difference in the segmentations between our model and nnUNet is likely due to our improved training strategy involving cropping along the center of the spinal cord, regression-based adaptive wing loss, and most importantly, training \textit{directly} on the soft GT masks (unlike in nnUNet where the soft GT were binarized). The localization of the spinal cord, mainly through cropping, has been a recurrent prerequisite step in the literature \citep{Gros2019-uq, Benjdira2020-ah, LEMAY2021102766}, suggesting its necessity for obtaining robust segmentations. 

Furthermore, our model \texttt{soft\_all} is able to better delineate the shape of the spinal cord even in the presence of lesions compared to models trained specifically on one contrast (\texttt{soft\_per\_contrast T2w} and DeepSeg2D). This enhanced performance can likely be attributed to the model's comprehensive exposure to diverse contrasts of the spinal cord, cerebrospinal fluid, gray matter and white matter, that are featured in the Spine Generic \citep{Cohen-Adad2021-su} dataset. 

\subsection{Limitations \& Future Work}

\subsubsection{Application to thoracic and lumbar levels}
The proposed model is trained on a dataset of healthy participants containing cervical and upper-thoracic spinal cords only. Future research will add images with lumbar cords to further improve the generalizability of the model towards different fields-of-view.

\subsubsection{Center cropping}
The center cropping step in the online preprocessing of the images during training and inference assumes that spinal cord is at the center of the image. While this is a reasonable assumption, there might be some edge cases containing lumbar spines or participants with scoliosis on which the automatic predictions might fail. In order to mitigate this issue, the SCT function that runs inference with the proposed model 
provides a flag to change the default cropping (allowing researchers to adjust the cropping sizes based on their images).

\textblue{
\subsubsection{Binarization threshold}
At prediction time, the soft output was binarized for comparison with other methods. That binarization was done using a 0.5 threshold. As discussed in \citep{Gros2021-ms}, that threshold could potentially be optimized to further reduce the variability of CSA across contrasts. However, this would imply arbitrarily categorizing images into a given contrast, which defeats the purpose of the current contrast-agnostic method, wherein the model can be used as is regardless of the acquisition parameters. Moreover, with MRI acquisitions, it is difficult to define the contrasts accurately as, for instance, some combinations of parameters could lead to more/less T2w or more/less magnetization transfer saturation depending on the offset of the MT pulse and/or the presence of saturation bands. 
}

\textblue{\subsubsection{Validation in pathologies}}
\textblue{As mentioned in the introduction, one of the advantages of soft segmentation is the ability to encode volumetric measures with finer precision compared to binary segmentation. Considering that changes in the spinal cord happen at a very slow rate at early stages of MS, we expect that soft segmentation of the spinal cord will help detect subtle atrophies. We expect that the soft segmentation of the spinal cord will increase the precision of spinal cord CSA measurements, and thus lead to lower arm-size in trials \citep{Bautin2021-qe, Cawley2018-ht} . This will especially be of interest in the context of cross-sectional studies where protocols can vary. In patients with degenerative cervical myelopathy, which is characterized by a progressive compression of the spinal cord, being able to precisely segment the spinal cord could also lead to better prognostication and therapeutic strategies ~\cite{Martin2018-qt}.} 

\subsubsection{Continual model refinement through active learning}
The improvement in the mitigation of biases in morphometric measures between MRI contrasts holds exciting prospects for future work. Thanks to the remarkable generalization of the proposed model to unseen contrasts, datasets containing other contrasts (e.g., phase-sensitive inversion recovery, short tau inversion recovery, susceptibility-weighted imaging, MP2RAGE) can be added to the existing cohort to further improve the generalizability of the model. Moreover, enriching the model with images from patients with spinal pathologies will likely improve zero-shot generalization on participants with lesions and/or spinal cord compressions. 
For these advancements, human-in-the-loop active learning \citep{BUDD2021102062} involving an initial batch of segmentations from our model followed by manual corrections of under-/over-segmentations and then re-training the model on the larger datasets until fully automatic predictions are obtained is an attractive strategy. This approach for gradually aggregating large-scale datasets while improving the model simultaneously is similar to the recently proposed Segment Anything \citep{kirillov2023segany} model, thus paving way towards a foundational model for contrast-agnostic spinal cord segmentation.  

\section{Conclusion}
\label{sec:conclusion}
We presented an automatic method for the soft segmentation of the spinal cord across various MRI contrasts. Using a novel framework for generating soft GT masks that leverages information from all contrasts along with aggressive data augmentation and a regression loss function, our model produces soft segmentations that encode partial volume information at the spinal cord and cerebrospinal fluid boundary \textblue{and produces segmentations that are stable across contrasts}. More importantly, the proposed model reduces variability in CSA across all contrasts and generalizes well to unseen contrasts, pathologies, and MRI vendors. Overall, the proposed method could potentially increase the sensitivity of CSA as a biomarker to detect subtle spinal cord atrophy~\cite{Bautin2021-qe}.

\section*{Acknowledgments}
We thank Étienne Bergeron, Olivier Lupien-Morin, Benjamin Carrier and Yassine El Bouchaibi for helping with the manual
corrections and labeling, Nicholas Guenther, Alexandru Foias and Mathieu Guay-Paquet for data management with git-
annex. We thank Konstantinos Nasiotis, Charley Gros, Uzay Macar, Adrian El Baz and Jan Valošek, Louis-François Bouchard for the insightful discussions. 

This study was funded by funded by the Chair in Quantitative Magnetic Resonance Imaging [950-230815], CIHR [CIHR FDN-143263], the Canada Foundation for Innovation [32454, 34824], Fonds de Recherche du Québec - Santé [322736], NSERC of Canada [RGPIN-2019-07244], Canada First Research Excellence Fund, Courtois NeuroMod project, Quebec BioImaging Network [5886, 35450], INSPIRED (Spinal Research, UK; Wings for Life, Austria; Craig H. Neilsen Foundation, USA), Mila - Tech Transfer
Funding Program, and National Institute of Neurological Disorders and Stroke of the National Institutes of Health (USA) [K23NS104211 and L30NS108301]. ENK is supported by the Fonds de Recherche du Quebec Nature et Technologies B2X Doctoral scholarship and UNIQUE Excellence Doctoral scholarship.  SB is supported by the Fonds de Recherche du Quebec Nature et Technologies master’s training scholarship and by the Natural Sciences and Engineering Research Council of Canada graduate scholarship.

\bibliographystyle{elsarticle-harv.bst}\biboptions{authoryear}
\bibliography{paperpile,paperpile-naga,traditional-bib}

\clearpage

\twocolumn[
    \begin{@twocolumnfalse}
        \begin{center}
            \large\textbf{Supplementary Material: Towards contrast-agnostic soft segmentation of the spinal cord}
        \end{center}
    \end{@twocolumnfalse}
]

\setcounter{section}{0}
\renewcommand{\thesection}{\arabic{section}}
\section{Supplementary material}


\begin{suppfigure*}[t!]
\centering
\includegraphics[width=\linewidth]{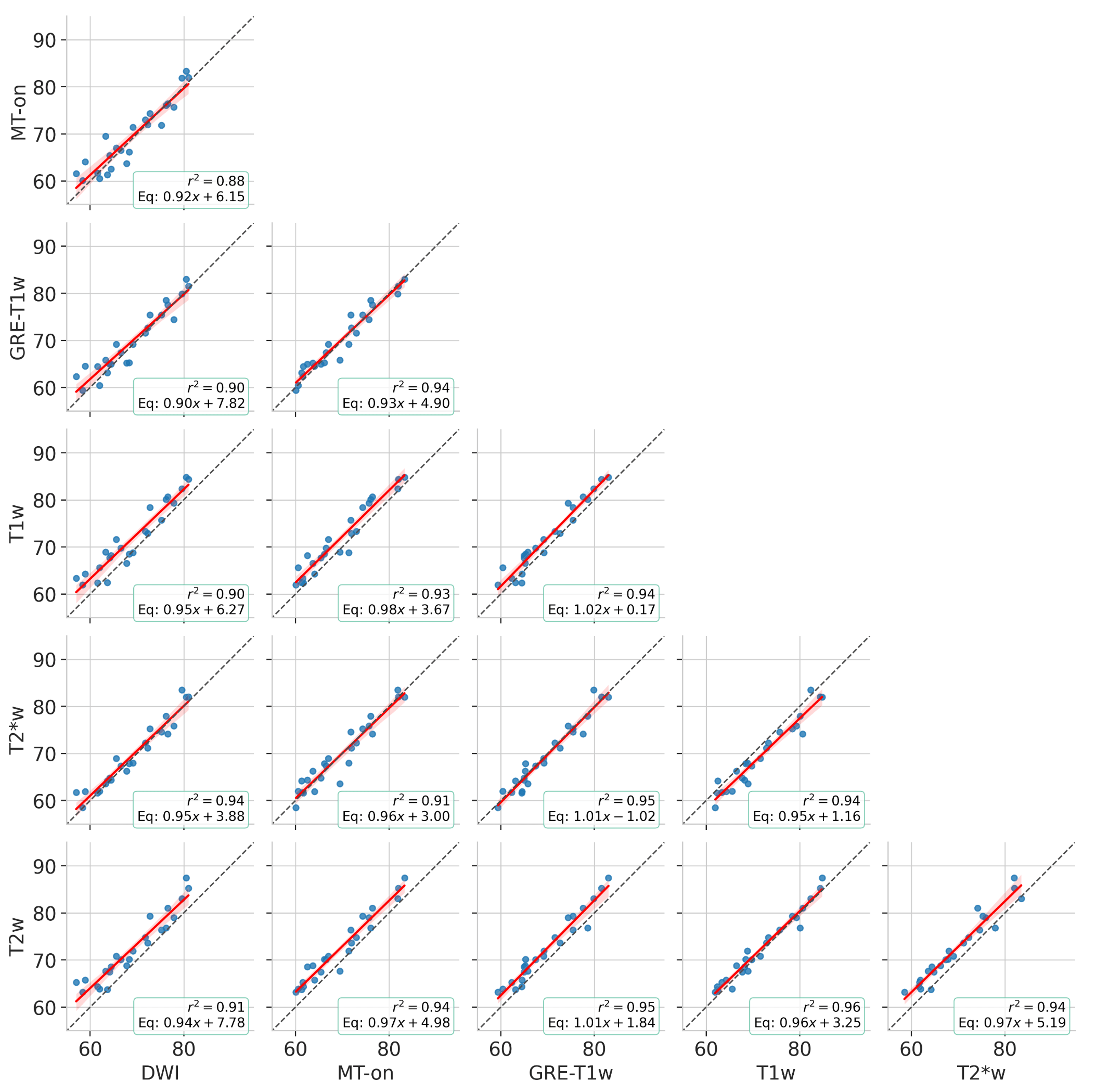}
\suppcaption{Pairwise correlation plots showing the level of agreement between CSA for each pair of contrasts for the proposed \texttt{soft\_all} model. Each scatter point represents one participant and the dashed line corresponds perfect agreement.}
\label{fig:pairwise-all}
\end{suppfigure*}

\subsection{Pairwise correlation plots for all six MR contrasts}

\Cref{fig:pairwise-all} shows the $r^2$ correlation plot between CSA at C2-C3 vertebral levels for each pair of contrasts used for the proposed \texttt{soft\_all} model. We observe a high level of agreement between MT-on / DWI (row 1, column 1), GRE-T1w / MT-on contrasts (row 2, column 2), T2*w / MT-on (row 4, column 2), T2*w / GRE-T1w (row 4, column 3), and T2w / T1w contrasts (row 5, column 4). 

\subsection{Baselines: Comparison between absolute CSA error per contrast}

\Cref{fig:hard-all-abs-csa-error-per-contrast,fig:soft-per-contrast-abs-csa-error-per-contrast,fig:soft-all-diceCE-loss-abs-csa-error,fig:hard-all-diceCE-loss-abs-csa-error-per-contrast} show the absolute CSA error across MRI contrasts for each baseline, except for the error plot of the proposed \texttt{soft\_all} model which is presented in Figure 3 of the main manuscript. 

\subsubsection{Soft vs. Hard ground-truth masks}
Comparing \textblue{\texttt{hard\_all\_SoftSeg}} (\Cref{fig:hard-all-abs-csa-error-per-contrast}) and \texttt{soft\_all} (Figure 3), the difference between the two methods can be explained by the fact that \textblue{\texttt{hard\_all\_SoftSeg}} was trained with binary (hard) GT masks, whereas \texttt{soft\_all} was trained with averaged soft GT masks, \textblue{both using the adaptive wing loss}. We notice larger absolute CSA errors for the \textblue{\texttt{hard\_all\_SoftSeg}} model, especially for the T2*w, DWI and MT-on contrasts. Considering that these 3 contrasts are acquired with thick axial slices (and hence suffer from higher partial volume compared to the T1w and T2w), it is likely that the discrepancy between these models highlights the capability of our proposed procedure for the creation of soft GT masks for training that better encode partial volume information. Note that the DWI contrast is an average across diffusion directions after motion correction. Despite the fact that motion correction is applied, slight residual motion across volumes blurs the edges of the spinal cord when averaging the volumes, resulting in higher partial volume. This could also explain the higher CSA error of \textblue{\texttt{hard\_all\_SoftSeg}} compared to \texttt{soft\_all} for DWI contrasts.

\begin{suppfigure}[t!]
\centering
\includegraphics[width=.475\textwidth]{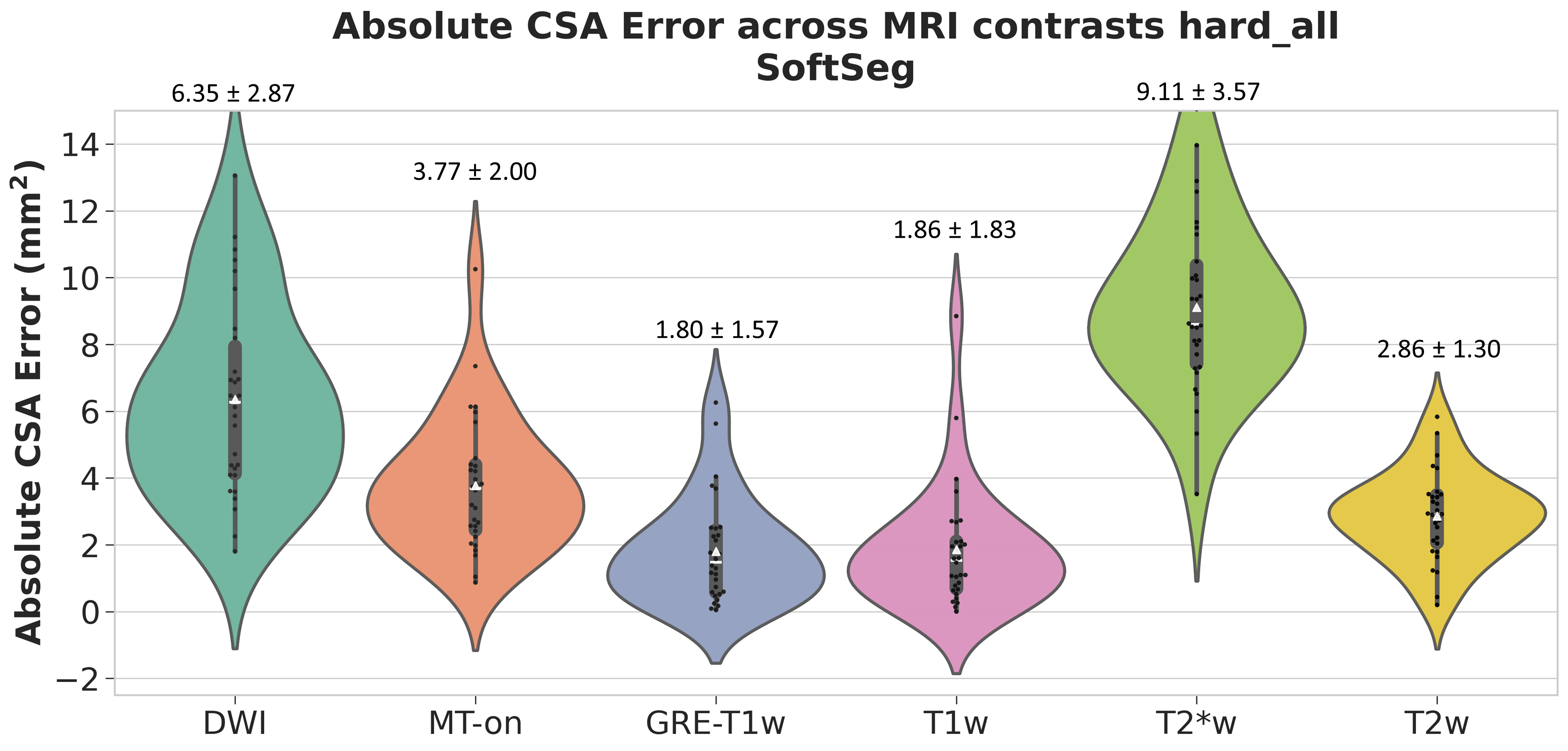}
\suppcaption{Absolute CSA error between the predictions and GT across each contrast for the \texttt{\textblue{hard\_all\_SoftSeg}} model trained on all contrasts with hard GT masks. Scatter plots within each violin represent the individual CSA errors for all test participants. White triangle marker shows the mean CSA error.}
\label{fig:hard-all-abs-csa-error-per-contrast}
\end{suppfigure}

\subsubsection{Single model vs. contrast-specific models}

\Cref{fig:soft-per-contrast-abs-csa-error-per-contrast} shows the absolute CSA errors when an individual model is trained for each of the 6 contrasts. Out of all the baselines, this achieves similar results compared to the \texttt{soft\_all} model, albeit with higher errors. Training independent models on each contrast does not help mitigating CSA variability, as they optimize for spinal cord segmentation separately, resulting in different CSA for a given participant, even when using soft GT masks. On the other hand, training a single model on all contrasts (i.e. \texttt{soft\_all}) simultaneously accounts for the heterogeneity in the appearance of images, thus improving CSA estimation. It is also useful to note that the \texttt{soft\_per\_contrast} models are trained on $6 \times $ less data compared to the single \texttt{soft\_all} model.

\begin{suppfigure}[htbp!]
\centering
\includegraphics[width=.475\textwidth]{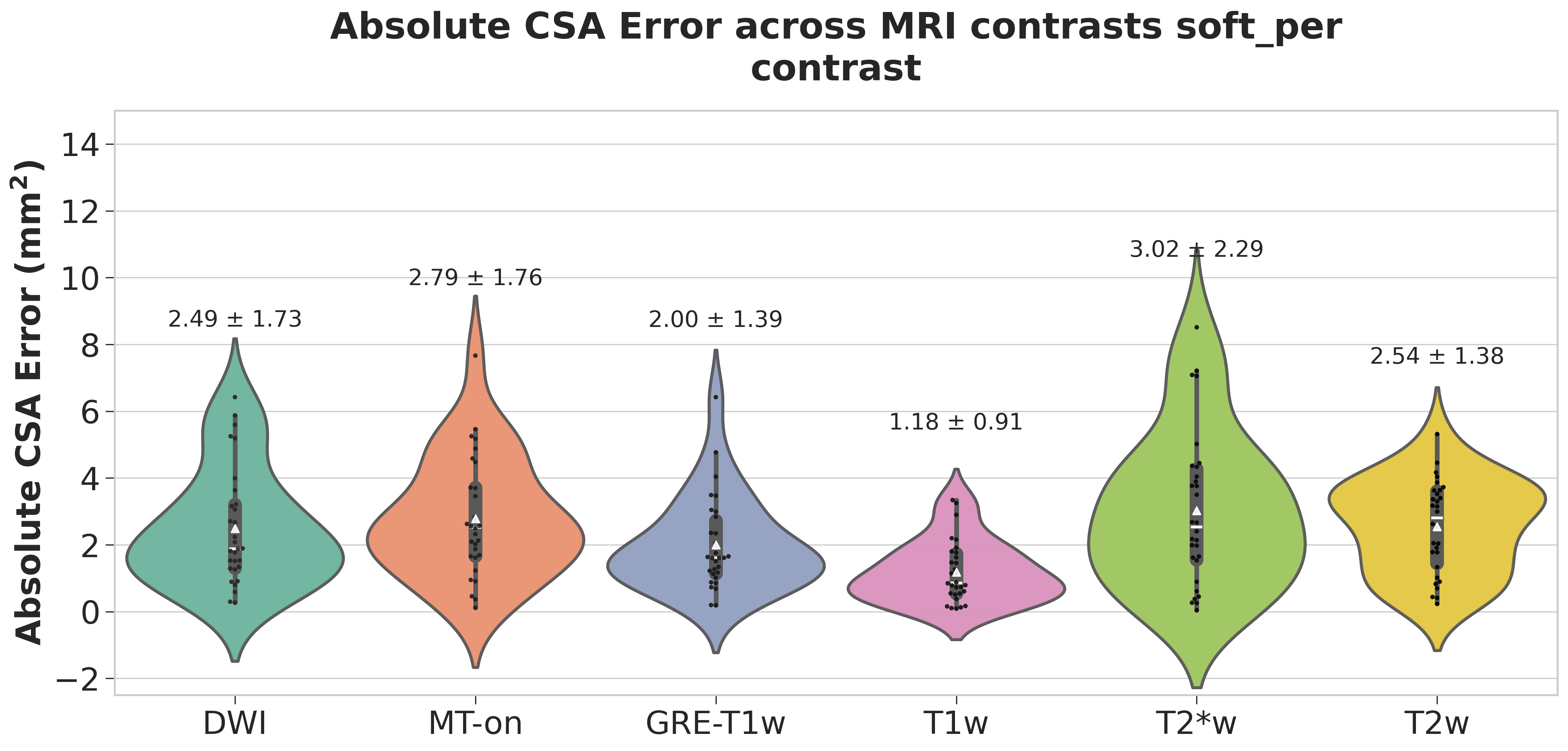}
\caption{Absolute CSA error between the predictions and GT across each contrast for the six models trained on each contrast independently with soft GT masks. Scatter plots within each violin represent the individual CSA errors for all test participants. White triangle marker shows the mean CSA error.}
\label{fig:soft-per-contrast-abs-csa-error-per-contrast}
\end{suppfigure}
 

\subsubsection{Dice cross-entropy vs. adaptive wing loss}
As mentioned in Section 2.3.4 of the main manuscript, using Dice coefficient in the training objective does not optimize for the accuracy of the segmentations at the spinal cord/cerebrospinal fluid boundary. This led to subtle under-segmentations across participants, thereby resulting in consistently larger absolute CSA errors by more than an order of 1 $\textrm{mm}^2$ across all contrasts. \Cref{fig:hard-all-diceCE-loss-abs-csa-error-per-contrast,fig:soft-all-diceCE-loss-abs-csa-error} show the CSA errors per contrasts for models trained with the \textblue{DiceCE} loss using hard masks and averaged soft masks, respectively. As expected, for the \texttt{hard\_all\_diceCE\_loss} model, the individual CSA estimations across contrasts vary substantially. For the \texttt{soft\_all\_diceCE\_loss} model, we see a relative improvement across contrasts, but does not outperform the \texttt{soft\_all} model shown in Figure 3 of the main paper.

\begin{suppfigure}[t!]
\centering
\includegraphics[width=.475\textwidth]{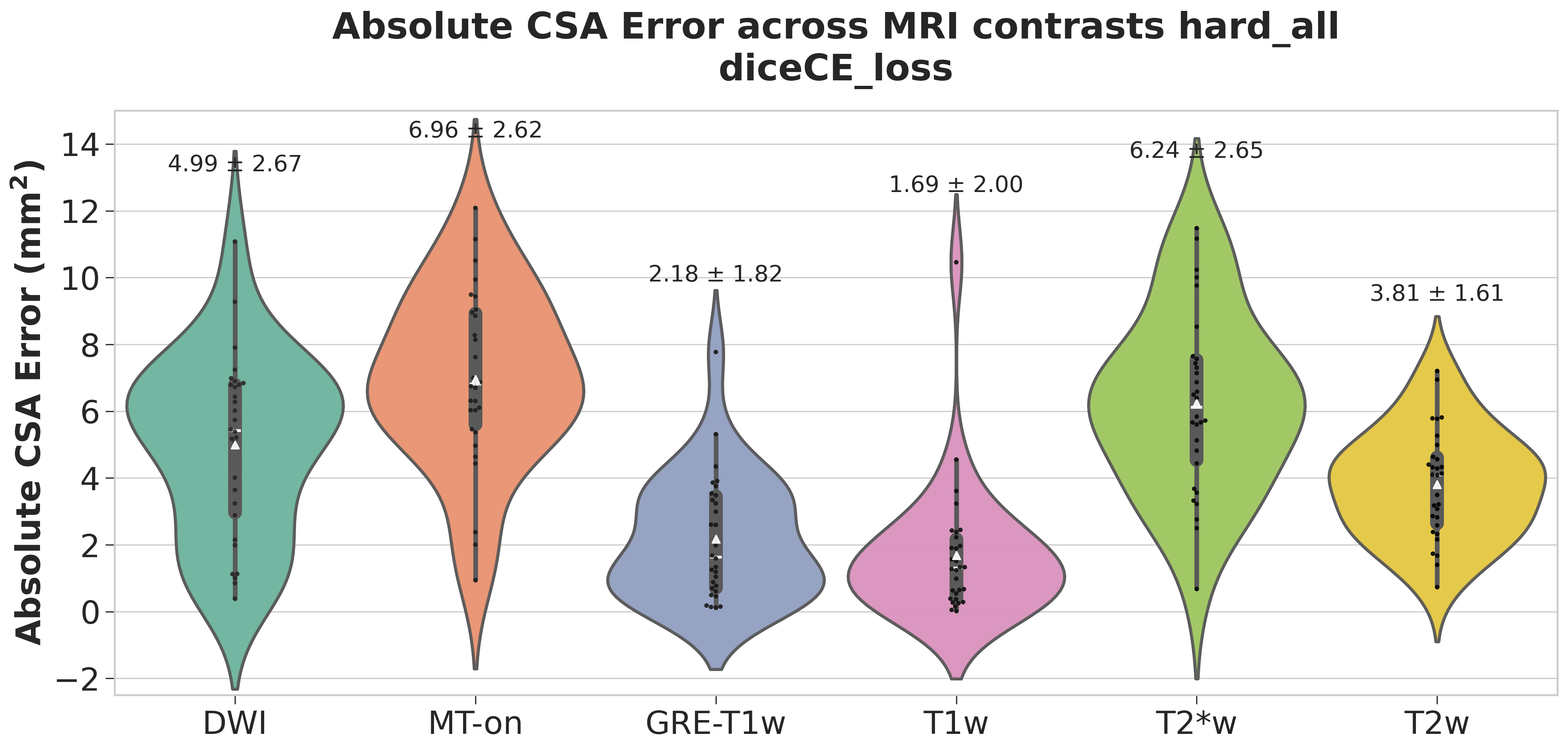}
\suppcaption{\textblue{Absolute CSA error between the predictions and GT across each contrast for the model trained on all contrasts with hard GT masks and Dice cross-entropy loss (instead of adaptive wing loss). Scatter plots within each violin represent the individual CSA errors for all test participants. White triangle marker shows the mean CSA error.}}
\label{fig:hard-all-diceCE-loss-abs-csa-error-per-contrast}
\end{suppfigure}

\begin{suppfigure}[htbp!]
\centering
\includegraphics[width=.475\textwidth]{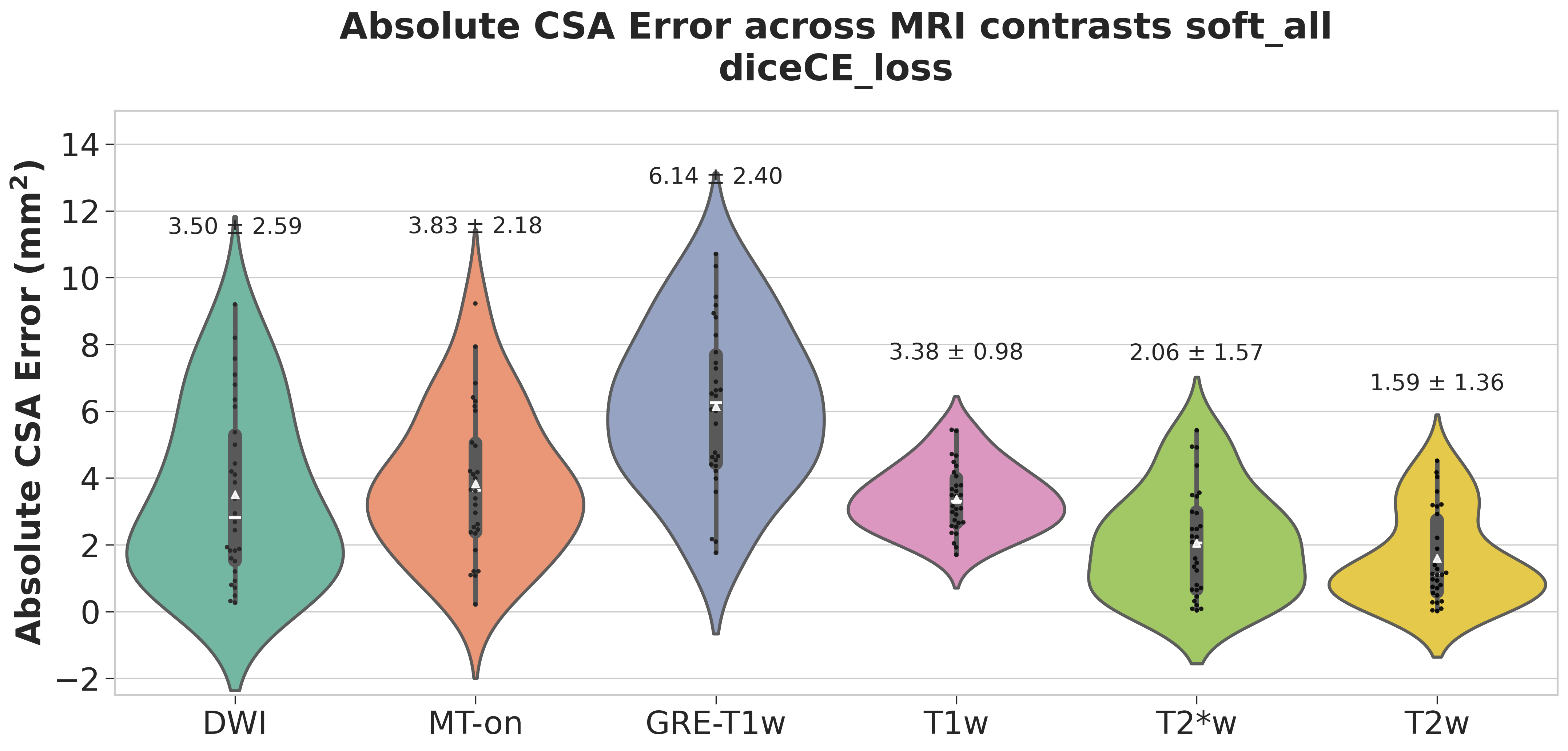}
\suppcaption{\textblue{Absolute CSA error between the predictions and GT across each contrast for the model trained on all contrasts with soft GT masks and Dice cross-entropy loss (instead of adaptive wing loss). Scatter plots within each violin represent the individual CSA errors for all test participants. White triangle marker shows the mean CSA error.}}
\label{fig:soft-all-diceCE-loss-abs-csa-error}
\end{suppfigure}

\subsection{State of the art: Comparison between absolute CSA error per contrast}

\Cref{fig:deepseg2d-abs-csa-error-per-contrast,fig:hard-all-bigaug-abs-csa-error-per-contrast,fig:nnunet-abs-csa-error-per-contrast} show the absolute CSA error across MRI contrasts for the state-of-the-art methods. For comparison, the error plot for the proposed \texttt{soft\_all} model is presented in Figure 3 of the main manuscript.

Between DeepSeg2D and the proposed \texttt{soft\_all} model, we observe large errors especially for the DWI, MT-on, GRE-T1w, and T2*w contrasts. There are two possible reasons: (1) DeepSeg was not trained MT-on, and GRE-T1w contrasts, hence showing poor generalization, and (2) as mentioned before, the thick-slice acquisition of these 4 contrasts results in higher partial volume effect at the spinal cord / cerebrospinal fluid boundary, which is exacerbated by the hard GT masks used for training. In addition, the DWI contrast is subject to higher partial volume due to the averaging across diffusion direction, thus resulting in higher CSA errors. 

\begin{suppfigure}[t!]
\centering
\includegraphics[width=.475\textwidth]{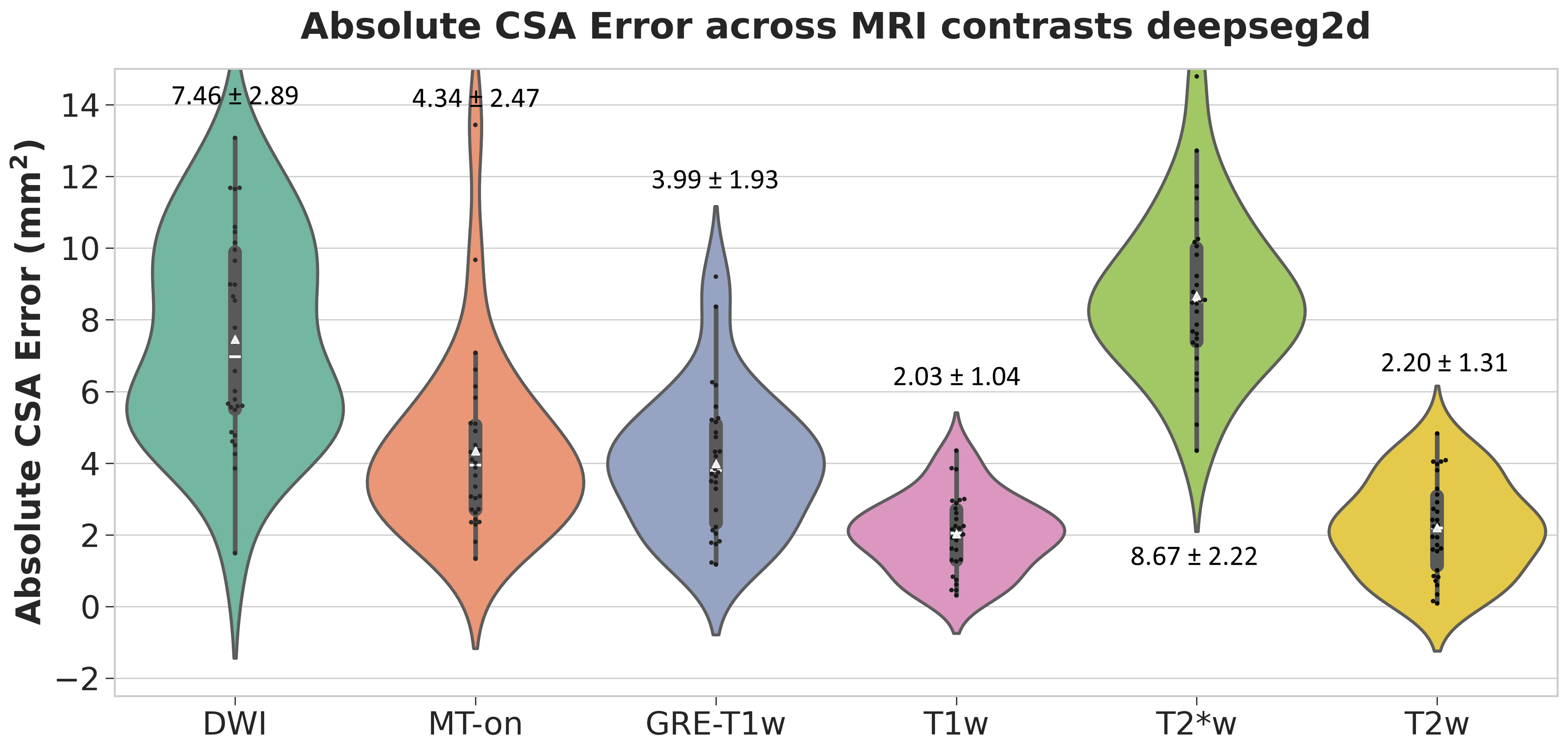}
\suppcaption{Absolute CSA error between the predictions and GT across each contrast for DeepSeg2D. Scatter plots within each violin represent the individual CSA errors for all test participants. White triangle marker shows the mean CSA error.}
\label{fig:deepseg2d-abs-csa-error-per-contrast}
\end{suppfigure}

Comparing \texttt{hard\_all\_BigAug} model's performance with our \texttt{soft\_all} model, we observed a similar trend in the variability of the CSA across contrasts as with the DeepSeg2D model. This is expected as both models were trained with individual contrast-specific hard GT masks.

\begin{suppfigure}[htbp!]
\centering
\includegraphics[width=.475\textwidth]{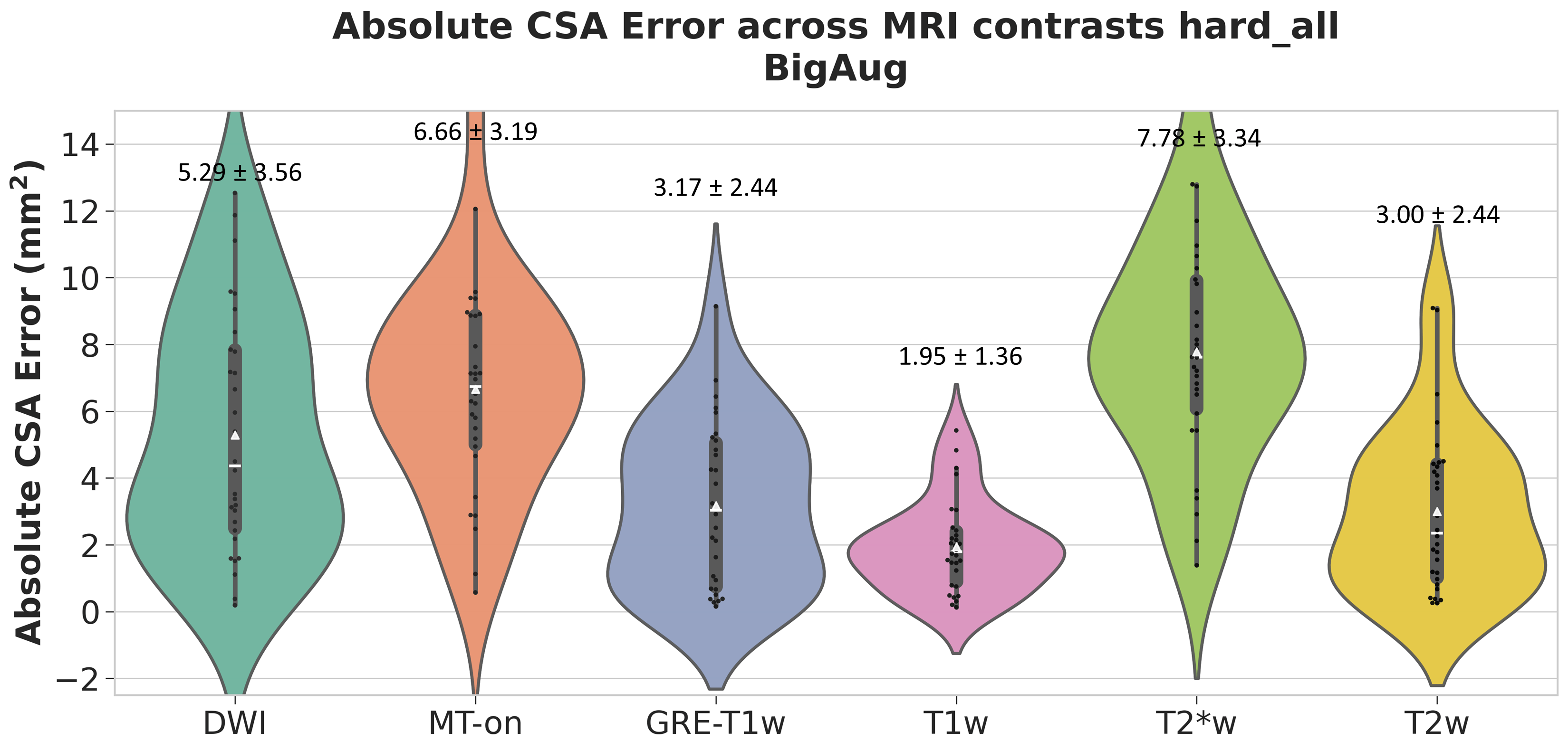}
\suppcaption{\textblue{Absolute CSA error between the predictions and GT across each contrast for \texttt{hard\_all\_BigAug}. Scatter plots within each violin represent the individual CSA errors for all test participants. White triangle marker shows the mean CSA error.}}
\label{fig:hard-all-bigaug-abs-csa-error-per-contrast}
\end{suppfigure}

\begin{suppfigure}[htbp!]
\centering
\includegraphics[width=.475\textwidth]{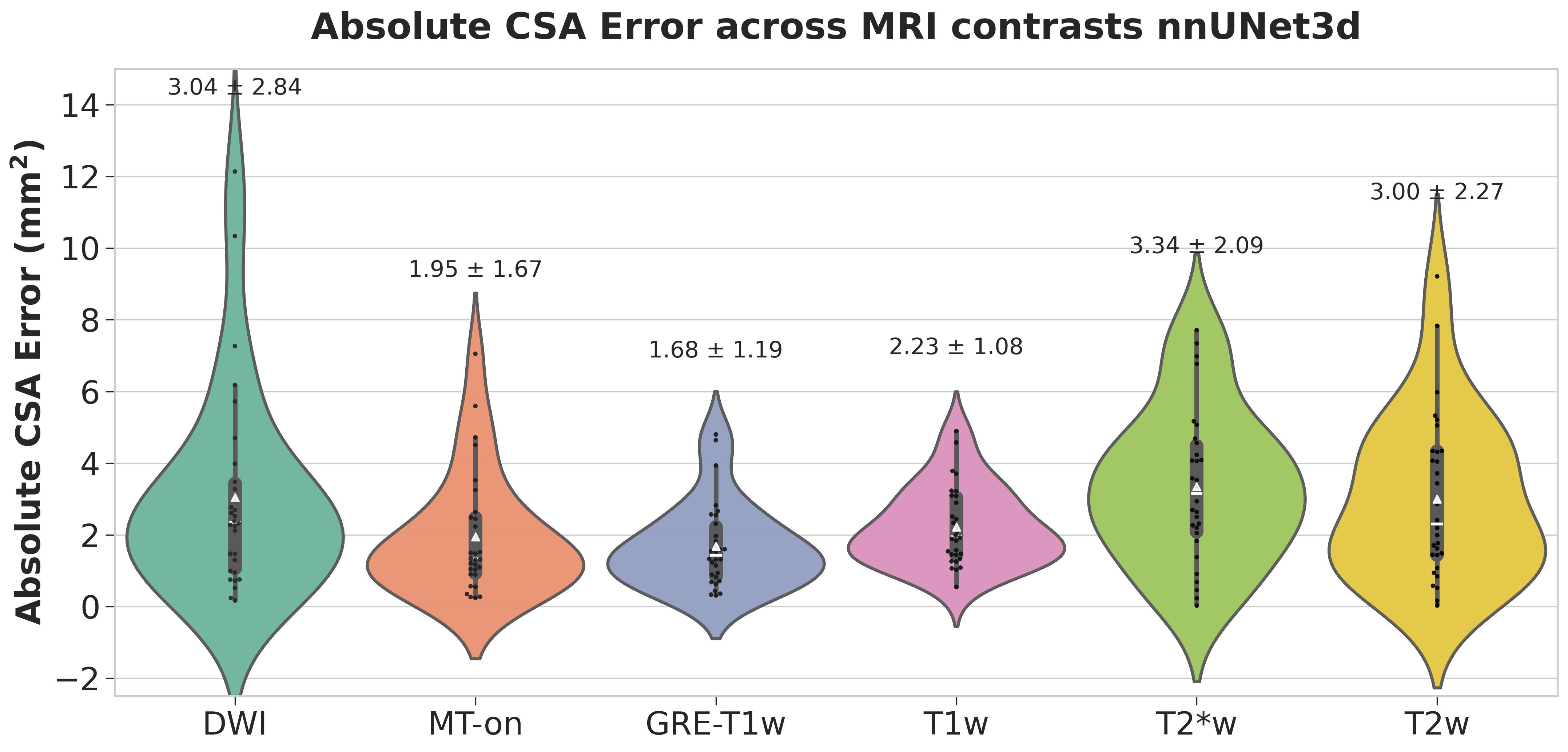}
\suppcaption{Absolute CSA error between the predictions and GT across each contrast for \textblue{nnUNet3D}. Scatter plots within each violin represent the individual CSA errors for all test participants. White triangle marker shows the mean CSA error.}
\label{fig:nnunet-abs-csa-error-per-contrast}
\end{suppfigure}

Comparing nnUNet's performance (\Cref{fig:nnunet-abs-csa-error-per-contrast}) with the \texttt{soft\_all} model, we notice higher CSA errors for DWI, T2*w, and T2w contrasts. While similar arguments from the DeepSeg models can be applied to explain the high DWI and T2*w errors, it is difficult to explain the higher error for T2w images. One possible reason is that nnUNet sets the patch size based on the median shape of all the images in the training set. As the training set is dominated by thick-slice acquisition images (4 out of 6 contrasts), this results in smaller median size with insufficient coverage of the spinal cord in the S-I direction (unlike the larger patch size that we choose for \texttt{soft\_all} via center-cropping), thus resulting in higher errors. 

\textblue{\subsection{State of the art: CSA variability across all methods}}

\textblue{
Figures \ref{fig:csa-std-all-methods} and \ref{fig:csa-error-all-methods} show the variability of CSA across \textit{all} methods. Note that subsets of these plots are reported in the main paper.}
Considering a comparison within the DeepSeg models, we see that the 2D model achieves relatively better results (i.e. lower CSA errors) than the 3D model. The worse performance of the 3D model can be explained by the \textit{patch size} chosen for inference with sliding windows. In the \href{https://github.com/spinalcordtoolbox/spinalcordtoolbox/blob/6.2/spinalcordtoolbox/deepseg_/sc.py#L344-L346}{source code}, we observed that the patch sizes were fixed to $64 \times 64 \times 48$ and $96 \times 96 \times 48$ depending on the contrast, which do not contain enough contextual information of the cord in the A-P and S-I axes. This means that these patch sizes unintentionally cut off patches of the cord, thereby not providing its complete structure. On the other hand, the 2D model uses individual slices in the S-I axes containing the complete cross-sectional view of the cord during inference. This results in a superior performance of the DeepSeg 2D compared to DeepSeg 3D. 

\textblue{
With the nnUNet models, nnUNet2D model performed slightly better in terms of the absolute CSA error (i.e. the error was slightly lower) and slightly worse standard deviation (STD) of cross-sectional area (CSA) across contrasts, compared to nnUNet3D. 
Within the nnUNet models, nnUNet3D used a patch size of 80 × 192 × 160 (RPI orientation) for training, while nnUNet2D used a patch size of 256 × 224 (PI orientation) slicing up the 3D volume along the R-L dimension. A larger patch-size, especially in the superior-inferior dimension (160 in 3D vs. 224 in 2D) resulted in the 2D model performing slightly better than the 3D model. Furthermore, the difference between nnUNet 2D and 3D is not as substantial as the one observed DeepSeg 2D and 3D because DeepSeg3D used a much smaller patch size ($64 \times 64 \times 48$ or $96 \times 96 \times 48$), failing to treat the (tubular) spinal cord structure as a whole. 
}

\textblue{
Lastly, nnUNet2D performed considerably better than DeepSeg2D mainly because it was trained on soft masks that were binarized at 0.5 threshold, further emphasizing that our proposed preprocessing pipeline for creating soft masks by combining multiple contrasts is key to reducing morphometric variability across contrasts. 
}

\begin{suppfigure*}[t!]
\centering
\includegraphics[width=0.95\textwidth]{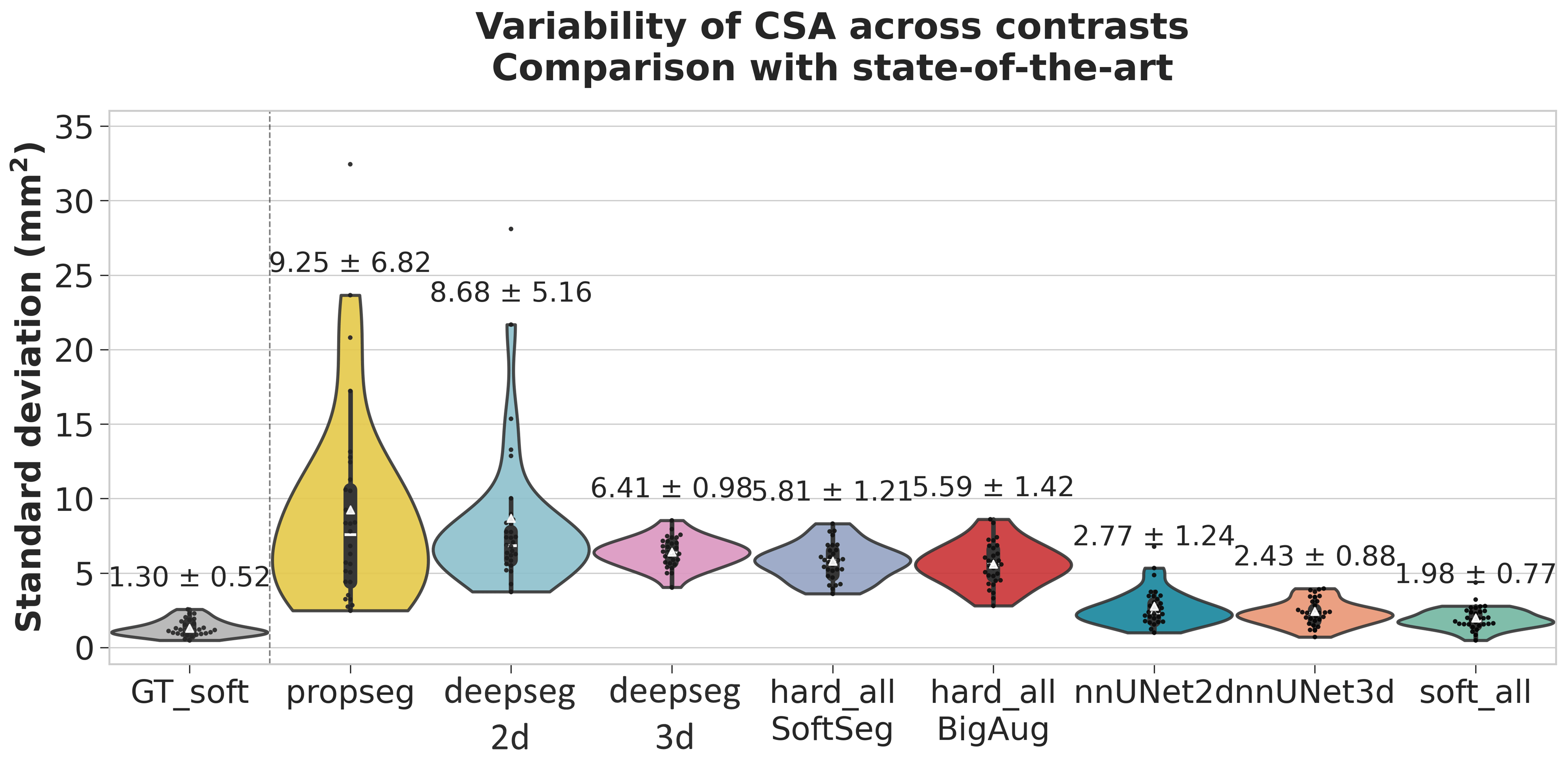}
\suppcaption{\textblue{Standard deviation of CSA between C2-C3 vertebral levels for
PropSeg, DeepSeg3D/2D, \texttt{hard\_all\_SoftSeg},  \texttt{hard\_all\_BigAug}, nnUNet3D/2D,  and our model \texttt{soft\_all}. White triangle marker shows
the mean CSA STD.}}
\label{fig:csa-std-all-methods}
\end{suppfigure*}

\begin{suppfigure*}[t!]
\centering
\includegraphics[width=.95\textwidth]{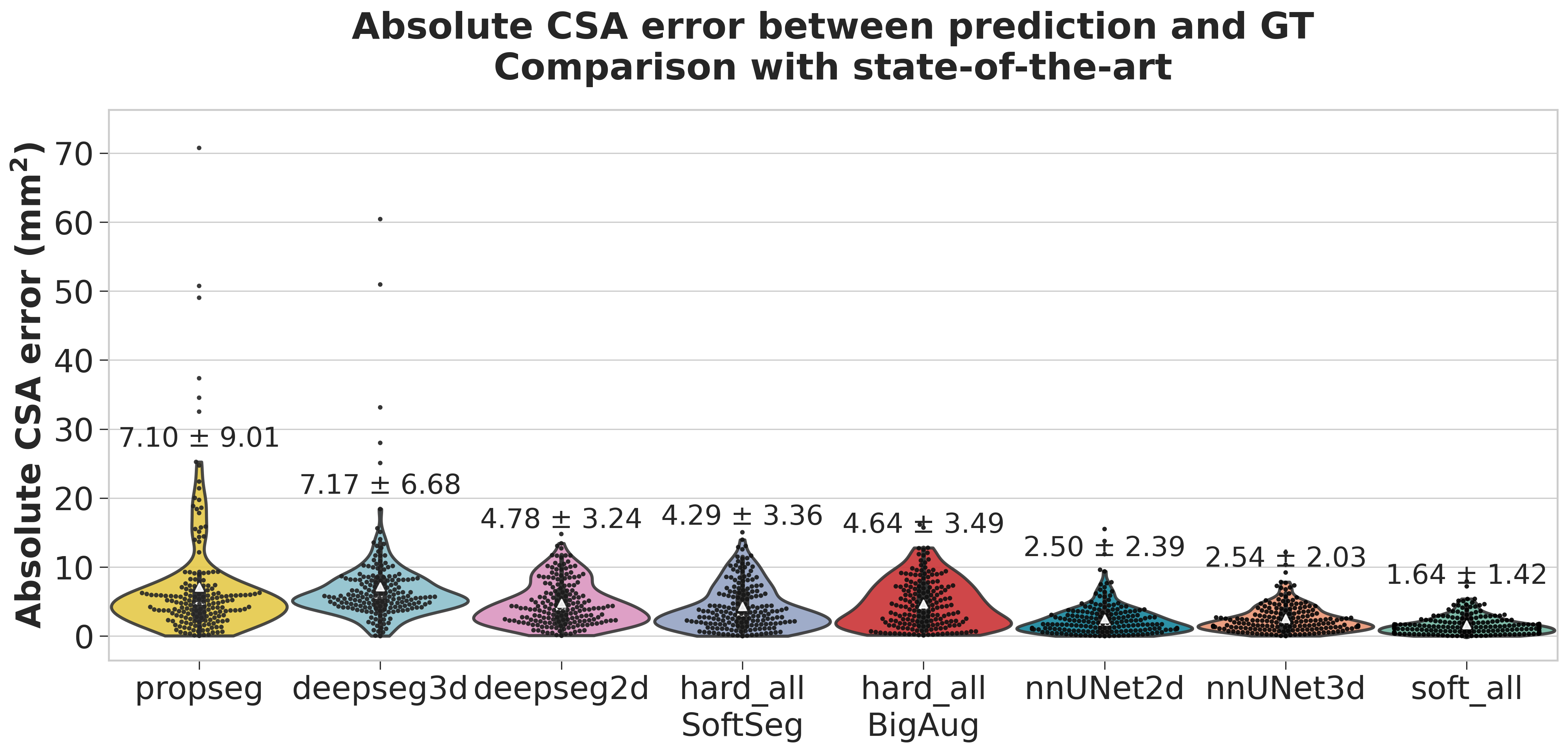}
\suppcaption{\textblue{Mean absolute CSA error for PropSeg, DeepSeg3D/2D,  \texttt{hard\_all\_SoftSeg},  \texttt{hard\_all\_BigAug}, nnUNet3D/2D, and our model \texttt{soft\_all}. White triangle marker showsthe mean CSA error.}}
\label{fig:csa-error-all-methods}
\end{suppfigure*}

\textblue{\section{Generalization to the FLAIR contrast}}

\begin{suppfigure}[htbp!]
\centering
\includegraphics[width=.475\textwidth]{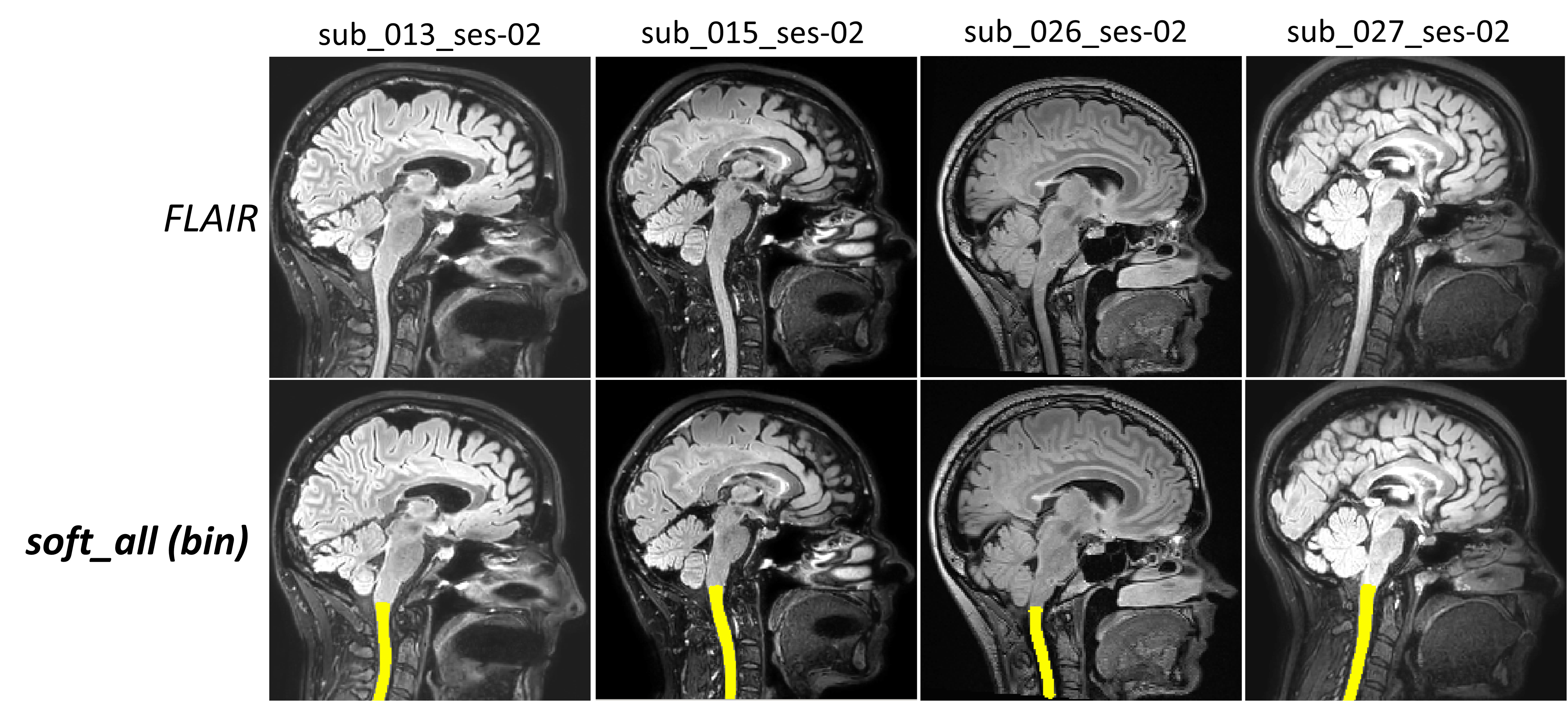}
\suppcaption{\textblue{Representative samples showing our model's zero-shot generalization on the FLAIR contrast.}}
\label{fig:flair}
\end{suppfigure}

\textblue{
In addition to the contrasts and pathologies presented in Section 2.4.4 of the main paper, we also tested our model on the FLAIR contrast. 
We used the publicly available MICCAI challenge dataset\footnote{\href{https://portal.fli-iam.irisa.fr/msseg-2/data/}{https://portal.fli-iam.irisa.fr/msseg-2/data/}} for the segmentation of new/modified brain MS lesions. The dataset consists of 3D FLAIR images of 100 patients with 2 timepoints acquired from 15 different MRI scanners (1.5T and 3T) among 3 vendors (Siemens, Philips, GE). No GT labels were provided for the spinal cord, hence we used this dataset to show zero-shot predictions on the FLAIR contrast. 
}

\textblue{
\section{Contrast-agnostic segmentation: Ablations across contrasts}}
\label{sec:ablation}
\textblue{
In this section, we performed two ablation studies to evaluate the stability of the model with respect to the number of contrasts used to train the model. For each ablation study, we preprocessed the data to generate a unique soft segmentation GT that only included the selected contrasts. The models were then trained using the same parameters as our \texttt{soft\_all} model for both ablations.
}

\textblue{
For the first ablation study, we trained the model with two contrasts only: T1w and T2w. These were chosen because of their large field-of-view (FOV) compared to the remaining contrasts. Panel A of \Cref{fig:ablation_csa,fig:ablation_error} respectively show the average CSA and CSA error across all contrasts from the test set. As the model was trained on the soft GT averaged from T1w and T2w, we expected to see little divergence in CSA for these two contrasts in the test set, which is indeed confirmed. Interestingly, the model also performed reasonably well on the MTon and GRE-T1w contrasts in terms of CSA estimate, which is likely due to the similar cord/CSF appearance between MTon and T2w and between GRE-T1w and T1w. That being said, the absolute mean error for MT-on (7.01 $mm^2$) and GRE-T1w (7.82 $mm^2$) is much larger than that for T1w (1.33 $mm^2$) and T2w (1.28 $mm^2$). However, the model clearly does not perform well for DWI and T2*w contrasts. 
}

\textblue{
For the second ablation study, we trained the model with four contrasts: T1w, T2w, DWI and T2*w. There, we expected results to be more favourable to the DWI and T2*w contrasts, which is indeed confirmed by \Cref{fig:ablation_csa,fig:ablation_error}, panel B. Overall, we observe that CSA values across contrasts (\Cref{fig:ablation_csa}B) are more similar across all 6 contrasts, even if only 4 were used for creating the GT masks and model training. It is important to note that the number of images in the training set doubled compared to the first ablation (T1w and T2w).
}

\begin{suppfigure}[htbp!]
\centering
\includegraphics[width=.475\textwidth]{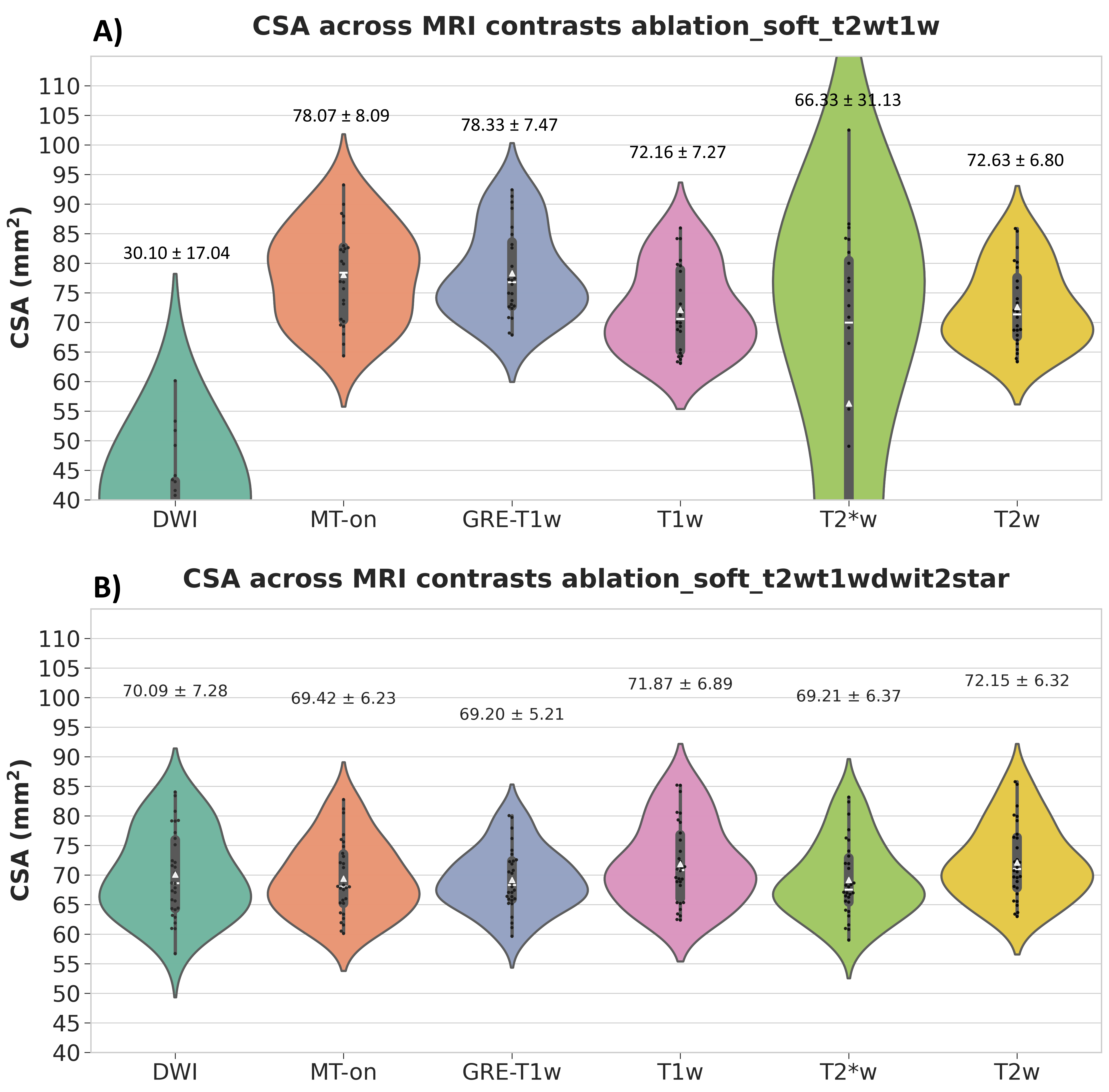}
\suppcaption{\textblue{Effect of number of contrasts included in the GT and training on CSA. A) CSA values of test set for a model that include T1w and T2w contrasts. B) CSA values of test set for a model that include T1w, T2w, DWI and T2*w contrasts. White triangle marker shows the mean CSA across participants.}}
\label{fig:ablation_csa}
\end{suppfigure}

\begin{suppfigure}[htbp!]
\centering
\includegraphics[width=.475\textwidth]{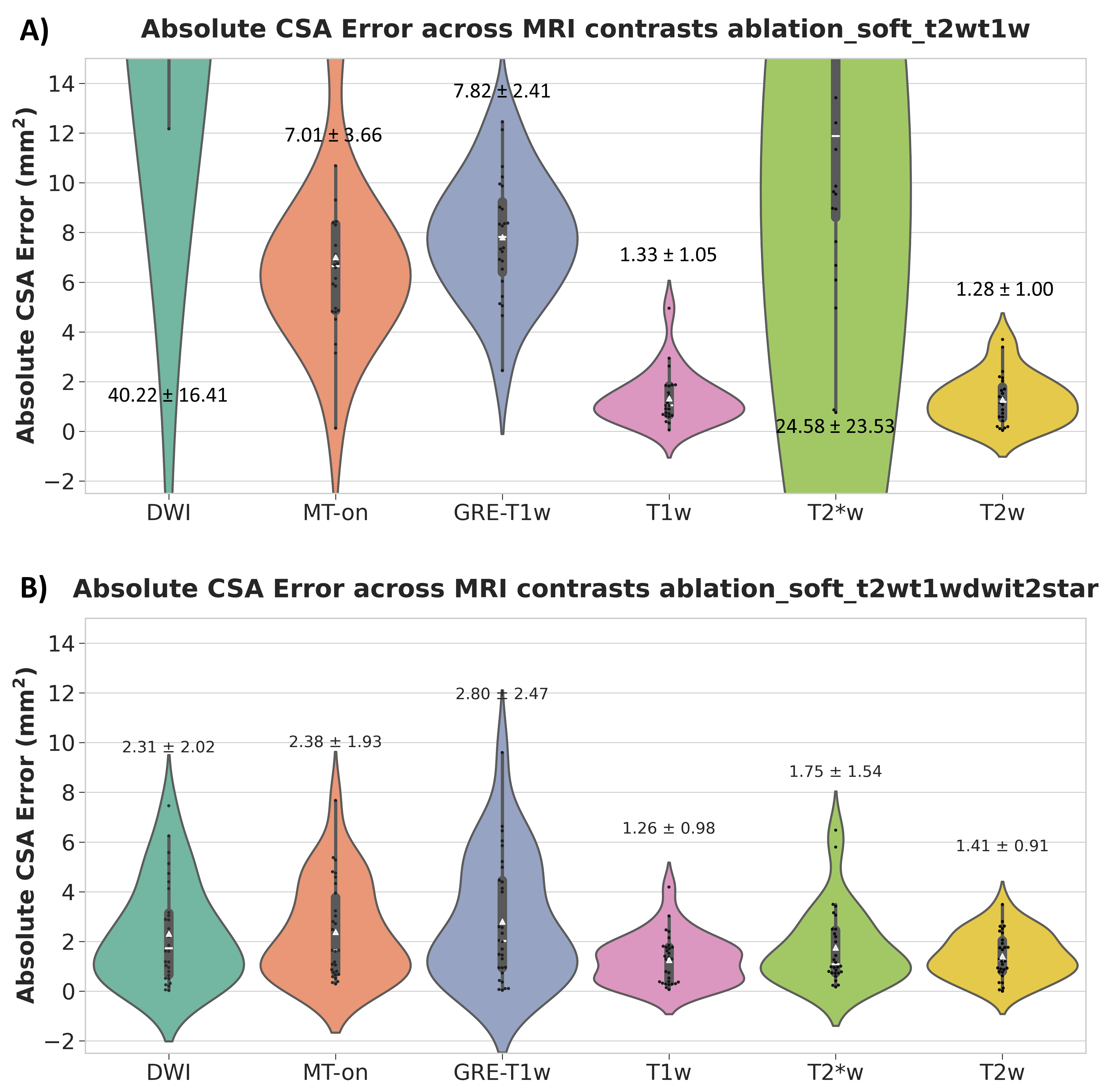}
\suppcaption{\textblue{Absolute CSA error between the predictions and GT for the \texttt{soft\_all} model including T1w and T2w contrasts (A) and for the \texttt{soft\_all} model including T1w, T2w, DWI and T2*w contrasts (B). Scatter plots within each violin represent the individual CSA errors for all participants in the test set. White triangle marker shows the mean CSA error across participants. }}
\label{fig:ablation_error}
\end{suppfigure}

\textblue{\section{SynthSeg for spinal cord segmentation}}
\textblue{
SynthSeg was originally proposed for the segmentation of brain scans of any resolution and contrast, however, it requires fully-labeled scans to synthesize brain images by sampling from a Gaussian Mixture Model (GMM). These synthetic images are then used to train the segmentation model. A notable challenge in re-training SynthSeg for spinal cord images is that it would require the labels for all the structures that are typically visible in spinal cord scans, such as the spinal cord, cerebrospinal fluid, vertebrae, bones, muscles, fat, lungs, heart, etc.
}

\textblue{
As it was out of the scope of this study to create labels for each anatomical region visible in the training dataset used here, we created the segmentations of four key structures only: (i) the spinal cord (spinal cord, which we already had), (ii) cerebrospinal fluid (CSF), (iii) vertebrae, and (iv) intervertebral discs. The labels for the three additional structures (ii-iv) were obtained using a preliminary version of the TotalSegmentatorMRI model\footnote{\href{https://github.com/neuropoly/totalspineseg}{https://github.com/neuropoly/totalspineseg}}. As our original dataset consists of cervical (and a few thoracic) spinal cord scans, the labels for the vertebrae and the discs were generated from C1-T12 and C1/C2-T11/T12, respectively and reoriented to RPI. The output classes for the segmentation model were defined as follows: $0: \textrm{spinal cord}$, $1: \textrm{CSF}$, $2: \textrm{Vertebrae}$, and $3: \textrm{Discs}$. This dataset comprised 209 labeled scans in total. 
}

\textblue{
For a fair comparison with our proposed model \texttt{soft\_all}, we updated the spatial deformation parameters, namely, flipping, shearing and bias field to lie close to the range with which our model was trained. The default activation function was also changed from \texttt{elu} to \texttt{relu}. The model was trained for 25 epochs with 1000 steps per epoch with a batch size of 2. 
\Cref{fig:synthseg} shows an output segmentation on the T1w contrast.
}

\begin{suppfigure}[htbp!]
\centering
\includegraphics[width=.4\textwidth]{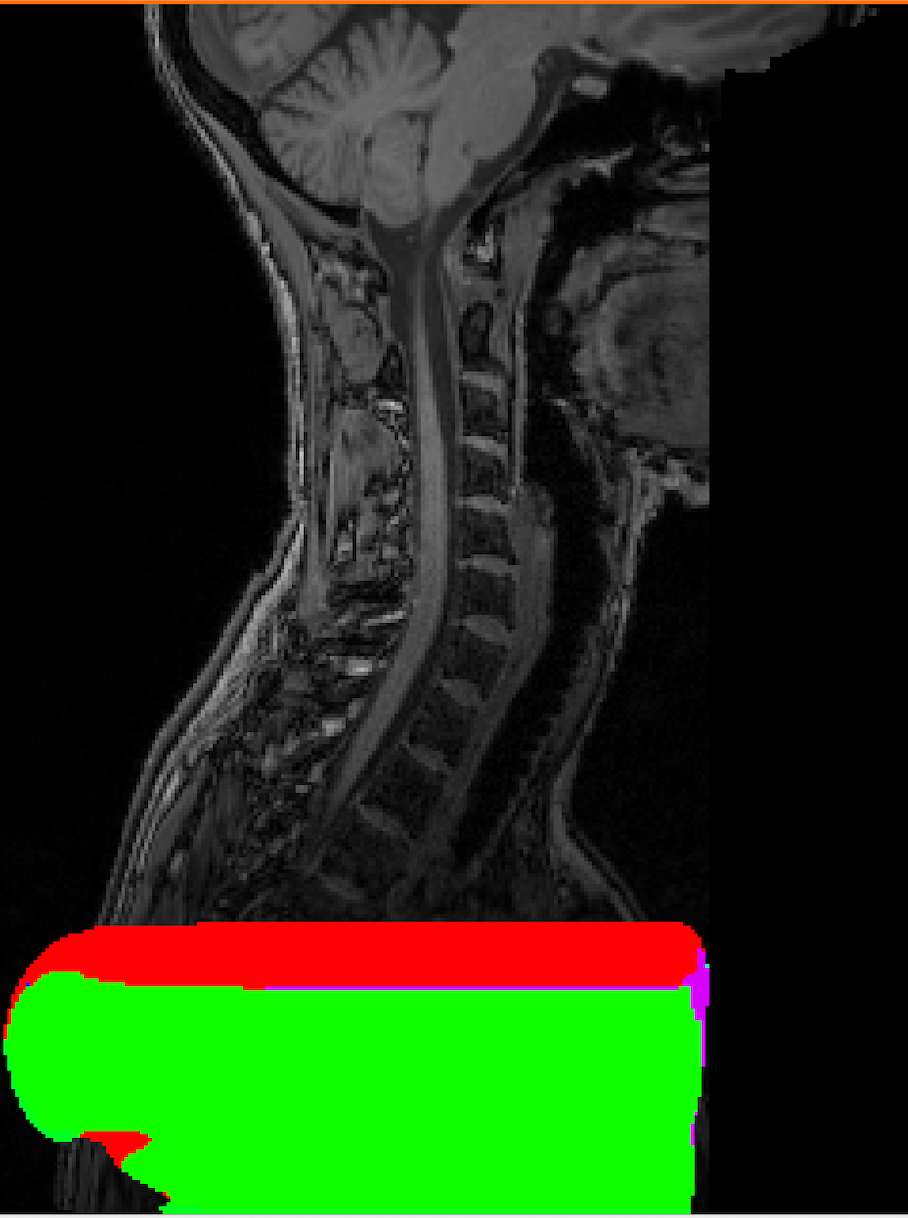}
\suppcaption{SynthSeg segmentation output on a T1w contrast of a healthy subject.}
\label{fig:synthseg}
\end{suppfigure}

\textblue{
We observed that the model struggled to segment the spinal cord structures and mostly output random segmentations. It must be noted that our dataset consisting of 209 scans is only a fraction compared to the 5000 brain scans used to train the original SynthSeg model. We performed several experiments tweaking various hyperparameters and the results were not comparable to the baseline methods described in the paper. 
More details on the re-training procedure along with a few synthetic scans and output segmentations can be found here\footnote{\href{https://github.com/sct-pipeline/contrast-agnostic-softseg-spinalcord/issues/111}{https://github.com/sct-pipeline/contrast-agnostic-softseg-spinalcord/issues/111}}.
}

\end{document}